\documentclass[aps,reprint,10pt,twocolumn,superscriptaddress]{revtex4-2}
\usepackage{amsmath, amssymb, amsthm, mathtools, bbm, physics}

\usepackage[utf8]{inputenc}
\usepackage{graphicx}
\usepackage{hyperref}

\usepackage{subcaption}
\usepackage{ragged2e} % for the \justifying macro
\DeclareCaptionJustification{justified}{\justifying}

\usepackage{microtype}
\microtypecontext{spacing=nonfrench}
\microtypesetup{
protrusion={true,nocompatibility},
activate={true,nocompatibility},
tracking=true,
kerning=true,
}

\usepackage{tikz} % TikZ
\usetikzlibrary{calc, arrows, patterns, shapes, decorations, arrows.meta, fit, positioning}
\tikzset{
    auto,node distance =1 cm and 1 cm,semithick,
    Var/.style ={circle, draw, minimum width = 1cm, ultra thick},
    Latent/.style ={regular polygon, regular polygon sides=3, inner sep=1pt, draw, minimum width = 1.3cm, ultra thick},
    var/.style ={circle, draw, minimum width = .5cm, ultra thick},
    latent/.style ={regular polygon, regular polygon sides=3, inner sep=1pt, draw, minimum width = .7cm, ultra thick},
    point/.style = {circle, draw, inner sep=0.06cm, fill, node contents={}},
    triangle/.style = {regular polygon, regular polygon sides=3, draw, inner sep=0.06cm, fill, node contents={}},
    bidir/.style={Latex-Latex,dashed},
    dir/.style={-Latex, thick},
    el/.style = {inner sep=2pt, align=left, sloped}
}
\tikzstyle{vertex}=[circle, fill=black!10, draw=black]
\tikzstyle{edge}=[thick]
\tikzstyle{clique}=[line width=4, draw=black!70]

\usepackage{pgfplots}
\pgfplotsset{compat=1.15}
\usepackage{mathrsfs}
\definecolor{ffbbqq}{rgb}{1,0.7333333333333333,0}
\definecolor{ffcctt}{rgb}{1,0.8,0.2}
\definecolor{qqqqcc}{rgb}{0,0,0.8}
\definecolor{zzccff}{rgb}{0.6,0.8,1}
\definecolor{rvwvcq}{rgb}{0.08235294117647059,0.396078431372549,0.7529411764705882}

\newtheorem*{defi}{Definition}

\newcommand{\mO}{\mathcal{O}}
\newcommand{\mL}{\mathcal{L}}

\newcommand{\Cset}{\mathcal{C}}
\newcommand{\Qset}{\mathcal{Q}}
\newcommand{\Nset}{\mathcal{N}}
\newcommand{\NBset}{\Nset(B)}

\newcommand{\comp}[1]{\overline{#1}}       
\graphicspath{ {./images/} }

\DeclareMathOperator{\Pa}{Pa}
\DeclareMathOperator{\pa}{pa}
\DeclareMathOperator{\Ch}{Ch}

\usepackage{xcolor}

\begin{document}
\title{Estimating the volumes of correlations sets in causal networks}

\author{Giulio Camillo}
\thanks{These authors have contributed equally to this manuscript. For correspondence: giulio.silva@usp.br.} 
%\email{giulio.silva@usp.br}

\affiliation{Instituto de Física, Universidade de São Paulo, 05508-220, São Paulo, Brasil}

\author{Pedro Lauand}  
\thanks{These authors have contributed equally to this manuscript. For correspondence: giulio.silva@usp.br.}
%\email{p223457@dac.unicamp.br}
\affiliation{Instituto de Física “Gleb Wataghin”, Universidade Estadual de Campinas, 130830-859, Campinas, Brazil}

\author{Davide Poderini}
\affiliation{International Institute of Physics, Federal University of Rio Grande do Norte, 59078-970, Natal, Brazil}

\author{Rafael Rabelo}
\affiliation{Instituto de Física “Gleb Wataghin”, Universidade Estadual de Campinas, 130830-859, Campinas, Brazil}

\author{Rafael Chaves}
\affiliation{International Institute of Physics, Federal University of Rio Grande do Norte, 59078-970, Natal, Brazil}
\affiliation{School of Science and Technology, Federal University of Rio Grande do Norte, Natal, Brazil}

\begin{abstract}
    Causal networks beyond that in the paradigmatic Bell's theorem can lead to new kinds and applications of non-classical behavior. Their study, however, has been hindered by the fact that they define a non-convex set of correlations and only very incomplete or approximated descriptions have been obtained so far, even for the simplest scenarios. Here, we take a different stance on the problem and consider the relative volume of classical or non-classical correlations a given network gives rise to. Among many other results, we show instances where the inflation technique, arguably the most disseminated tool in the community, is unable to detect a significant portion of the non-classical behaviors. Interestingly, we also show that the use of interventions, a central tool in causal inference, can enhance substantially our ability to witness non-classicality.
\end{abstract}

\maketitle

\section{Introduction}
The violation of Bell inequalities \cite{brunner2014bell} represents the strongest signature of non-classical behavior, as it can be observed in a device-independent context \cite{pironio2016focus}. Specifically, it demonstrates the incompatibility between quantum correlations and classical concepts of cause and effect, without relying on any assumptions about the internal mechanisms involved in the preparation and measurement of the physical system being analyzed. 

In the simplest Bell scenario, two distant parties measure two distinct dichotomic observables. However, this scenario has been extended and generalized in various ways. These extensions include incorporating additional measurements \cite{collins2004relevant} or expanding the number of possible outcomes \cite{collins2002bell}. Furthermore, the framework has been expanded to involve multiple parties \cite{werner2001all} and relaxations of locality assumptions \cite{pironio2003violations,brask2017bell}, as well as measurement independence \cite{hall2010local,chaves2015unifying}. Of particular significance are recent advancements that draw inspiration from causality theory \cite{pearl2009causality}. These generalizations explore networks with diverse topologies, growing in size and complexity, revealing a number of novel non-classical phenomena \cite{branciard2010characterizing,fritz2012beyond,renou2019genuine,weilenmann2020self,renou2021quantum,chaves2021causal,suprano2022experimental,polino2023experimental}.

In the context of a specific causal structure, a central inquiry arises: does it exhibit a classical-quantum gap? In other words, if the sources within the network are described by entangled quantum states, can measurements on them produce correlations that lack a classical interpretation? When considering networks with a single source, a classical depiction involves the characterization of a polytope \cite{pitowsky1991correlation}:  a convex set defined by a finite number of extremal points or, equivalently, a finite set of linear Bell inequalities.
However, even in such cases, the problem is recognized as intrinsically challenging, residing in the NP-hard complexity class \cite{pitowsky1991correlation}. This difficulty is further amplified when independent sources of correlations exist, as is often the case in paradigmatic quantum networks. In such scenarios, the correlations compatible with a given causal structure result in non-convex sets, necessitating computationally intensive algorithms rooted in algebraic geometry \cite{garcia2005algebraic} or various forms of approximation \cite{chaves2014proceedings,chaves2016polynomial,kela2019semidefinite,aaberg2020semidefinite,pozas2019bounding,wolfe2019inflation} proposed throughout the years. Not surprisingly, given these difficulties, there is still a very fragmented picture of the set of correlations causal networks can give rise to.

Here we propose an alternative route to gather further insights into the classical description of causal networks and their potential incompatibility with quantum correlations. Using quadratic optimization techniques (leveraging the Gurobi optimizer~\cite{gurobi}), already employed to address non-convex constraints originating from causal networks in~\cite{lauand2023witnessing}, as well as other tools such as the inflation technique~\cite{wolfe2019inflation}, the covariance approach~\cite{aaberg2020semidefinite} and Finner inequality~\cite{renou2019limits}, we calculate the volumes~\cite{cabello2005much,wolfe2012quantum} of classical and non-classical probability distributions that various causal structures can yield. As we show, in spite of its wide applicability and convergence in the asymptotic limit~\cite{navascues2020inflation}, in computational practice the inflation technique might offer relatively poor performance, since it is unable to reveal a significant portion of non-classical behaviors. Interestingly, similarly to what happens in the standard Bell scenario~\cite{duarte2018concentration}, we observe a concentration effect where most non-classical points concentrate an average distance from the classical set, the probability of finding points far away decaying exponentially. Finally, we show that the use of interventions~\cite{chaves2018quantum,gachechiladze2020quantifying} can significantly enhance our ability to detect non-classical behaviors.

The paper is organized as follows. In Sec.~\ref{sec:preli} we introduce the framework and methods underlying our work, including the definition of causal models and the three different causal networks that we study; we also describe the sampling method and the witness of non-classicality we have employed. In Sec.~\ref{sec:results}, for each of the three causal networks analyzed, we present the results on the volume estimation regarding different sets of correlations (classical, quantum, and non-signaling) as well as using different methods. In Sec.~\ref{sec:discussion} we discuss our findings and point out relevant questions and directions for future research.

\section{Preliminaries}
\label{sec:preli}

\subsection{Causal models}
% Decribe DAGs
% Classical, quantum, NS sets
The causal modeling framework~\cite{pearl2009causality,spirtes2000causation} offers a powerful language to describe causal constraints in terms of \emph{directed acyclic graphs} (DAG).
In this formalism, each node $A \in N(G)$ of the DAG $G$ is associated with a random variable and causal relationships are defined by the directed edges $E(G) \subseteq N(G) \times N(G)$ between these nodes.
In any real situation, we do not have access to every relevant cause that can influence our system, we then ought to distinguish between nodes that are associated with \emph{observable} variables $\mO_G \subseteq N(G)$ and unobserved, or \emph{latent}, ones $\mL_G \subseteq N(G)$.
Graphically, we will use circles and Latin letters for the former and triangles and Greek letters for the latter (see for example Fig.~\ref{fig:dags}).
Moreover, the random variables and the nodes will be represented with uppercase letters $A,B,\ldots$ while we will use the corresponding lowercase ones $a,b,\ldots$ to denote their outcomes.

Given a DAG $G$, we can define the concept of causal parents $\Pa(A)$ (or children $\Ch(A)$) of a given variable $A$ in $G$, as the set of nodes sharing incoming (or outgoing) edges with $A$.
This notion immediately give us a way to define what it means, for a classical joint distribution $p(\{a_i\}_i)$ on the random variables associated with observable nodes $A_i \in \mO_{G}$, to be compatible with a causal DAG $G$.

\begin{defi}[Classical compatibility]
A distribution $p(\{a_i\}_i)$ on the random variable associated with the nodes $\mO(G)$, is compatible with $G$ if it satisfies the following decomposition:
\begin{equation}
    p(\{a_i\}_i) = \sum_{\lambda \in \mL_{G}} \prod_{X \in N_{G}} p(x | \pa(X))
    \label{eq:markov_condition}
\end{equation}
where $\mL_{G} \subset N_{G}$ is the set of latent variables in $G$ and $\pa(X)$ the set of outcomes of all the parents of $X$.
\label{def:classical_set}
\end{defi}

The above decomposition is also called \emph{global Markov condition}.
We will use $\Cset(G)$ to represent the set of distributions compatible in this sense with a DAG $G$.

The compatibility notion defined above is valid only if we consider that our distribution $p$ arises from models where latent variables can be considered classical systems.
If instead, we allow them to be quantum systems, in general, we obtain a strictly larger set of compatible distributions, which we can denote by $\Qset(G)$ and for which generalizations of the global Markov condition and the concept of a causal structure have been proposed \cite{chaves2015information,pienaar2015graph,costa2016quantum,barrett2019quantum}

%NSI compatibility 
In some cases, it might be interesting to consider post-quantum resources distributed in the network \cite{henson2014theory}. The distributions that may arise from these models must respect some basic conditions that reflect the natural assumptions of no-signaling and independence of the sources (NSI). The core of the notion of the no-signaling principle to network scenarios is that the outcomes of one party should be insensitive to whatever the remaining parties do, including any local modifications in the particular arrangement of the topology of their part of the network. The set of correlations that arise from NSI has been explicitly studied primarily in the triangle scenario \cite{gisin2020constraints} and in the Evans scenario \cite{lauand2023witnessing}. We denote the set of correlations that is compatible with the principles of no-signaling and independence as $\mathcal{N}(G)$. We remark that the resulting constraints
derived from NSI will be valid for all general probabilistic theories \cite{PhysRevA.81.062348} (GPTs).

%\tmp{Add interventions... actually, are we going to talk about interventional data?}

\subsection{Single-source Bell scenario}

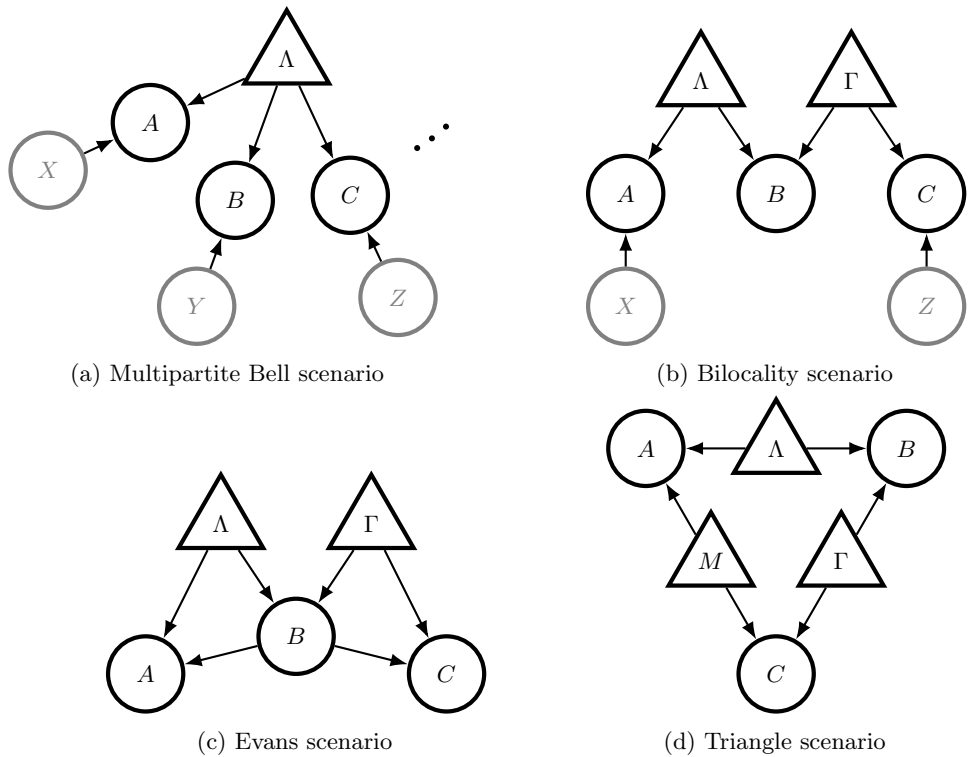
\begin{figure*}[t]
\begin{subfigure}[t]{.4\textwidth}
%\centering
    \begin{tikzpicture}  
        \node[Latent] (l) at (0,0) {$\Lambda$};
        \foreach \n/\a/\x in {1/A/X, 2/B/Y, 3/C/Z}{
            \draw (\n*180/4+160: 2cm) node[Var] (\a) {$\a$};
            \draw (\n*180/4+160: 3.5cm) node[Var, opacity=.5] (\x) {$\x$};
			\path[dir] (l) edge (\a) (\x) edge (\a);
        }
        \foreach \n in {1, 2, 3}{
            \fill (\n*17/3-40: 2cm + \n*0.1cm) circle[radius=0.04cm];
		}
    \end{tikzpicture}
    \caption{Multipartite Bell scenario}
    \label{fig:bell_dag}
\end{subfigure}
\begin{subfigure}[t]{.4\textwidth}
%\centering
    \begin{tikzpicture}
        \node[Latent] (l1) at (-1,1.5) {$\Lambda$};
        \node[Latent] (l2) at (1,1.5) {$\Gamma$};
        \node[Var] (a1) at (-2,0) {$A$};
        \node[Var] (b) at (0,0) {$B$};
        \node[Var] (a2) at (2,0) {$C$};
        \node[Var, opacity=.5] (x1) at (-2, -1.5) {$X$};
        \node[Var, opacity=.5] (x2) at (2, -1.5){$Z$};
        
        \path[dir] (x1) edge (a1) (x2) edge (a2);
        \path[dir] (l1) edge (a1) (l1) edge (b) (l2) edge (b) (l2) edge (a2);
    \end{tikzpicture}
    \caption{Bilocality scenario}
    \label{fig:bilocality_dag}
\end{subfigure}

\begin{subfigure}[t]{.4\textwidth}
%\centering
    \begin{tikzpicture}
        \node[Var] (a) at (-2,0) {$A$};
        \node[Var] (c) at (2,0) {$C$};
        %\node[var] (b) [right =of a] {$B$};
        \node[Latent] (l) at (-1,2) {$\Lambda$};
        \node[Latent] (g) at (1,2) {$\Gamma$};
        \node[Var] (b) at (0,.5) {$B$};
        % Directed edges
        \path[dir] (l) edge (a) (l) edge (b); 
        \path[dir] (g) edge (b) (g) edge (c); 
        \path[dir] (b) edge (a) (b) edge (c);
    \end{tikzpicture}
    \caption{Evans scenario}
    \label{fig:evans_dag}
\end{subfigure}
\begin{subfigure}[t]{.3\textwidth}
    \begin{tikzpicture}

        \foreach [count=\k] \l/\n/\a in {1/A/\Lambda, 2/C/M, 3/B/\Gamma} {
            \draw (\k*360/3 - 30: 1cm) node[Latent] (l\k) {$\a$};
            \draw (\k*360/3 + 30: 2cm) node[Var] (\n) {$\n$};
			\path[dir] (l\k) edge (\n);
        }
		\foreach \k/\l in {1/B, 2/A, 3/C}
			\path[dir] (l\k) edge (\l);

    \end{tikzpicture}
    \caption{Triangle scenario}
    \label{fig:triangle_dag}
\end{subfigure}
\caption{\textbf{Causal DAGs} \textbf{a)} Multipartite Bell scenario where the correlations between the distant parties are mediated by a single source of correlations. \textbf{b)} Bilocality scenario, akin to an entanglement swapping experiment \cite{pan1998experimental}, where two independent sources establish the correlations between three spatially separated parties. \textbf{c)} The Evans scenario with two independent sources of correlations but with a crucial difference to the bilocality case: the inputs for two of the parties are the measurement outputs of the central node, that is, the correlations are time-like and not space-like separated. \textbf{d)} The triangle scenario, where every party shares a bipartite and independent source of correlations with every other party.}
    \label{fig:dags}
\end{figure*}

Bell's paradigmatic causal structure (and multipartite generalization thereof) is composed by a number of parties, sharing classical correlations described by a random variable $\Lambda$, locally measuring different observables parametrized by $X,Y,Z, \ldots$ with corresponding measurement outcomes $A,B,C,\ldots$. This classical description is encoded in the class of DAGs $L_n$ represented in Fig.~\ref{fig:bell_dag}, implying that the observed distributions should follow the Markov condition given by 
\begin{eqnarray}\label{eq:classical_bell}
& & p(a,b,c,\cdots|x,y,z,\cdots) = \\ \nonumber
& & \sum_{\lambda} p(\lambda) p(a \vert x,\lambda)p(b \vert y,\lambda)p(c \vert z ,\lambda)\dots  .
\end{eqnarray}

%The first example was historically given in the notorious Bell scenario~\cite{}, that we can represent as a causal model using the DAG in Fig.~\ref{fig:bell_dag} specialized for two parties, where the incompatibility can be detected using the celebrated Bell inequalities, once the latent variable $\Lambda$ is used to represent a quantum state.

%In this case the nodes $A,B,C,\ldots$ are each associated with a set of POVM (positive operator valued measurements) described by operators $A^a_x, B^b_y, C^c_z, \ldots$, dependent on the value of the variables $X,Y,Z, \ldots$, while the node $\Lambda$ is associated with a multipartite state $\rho_\Lambda$.

In a quantum description, this probability distribution is given by the Born rule
\begin{equation}\label{eq:quantum_bell}
    p(a,b,c,\cdots|x,y,z,\cdots) = \tr \left[ \left( A^a_x \otimes B^b_y \otimes C^c_z \otimes \cdots\right) \; \rho_\Lambda \right] \, ,
\end{equation}
where the measurement inputs and outputs are associated with POVMs (positive operator valued measurements) and the classical node $\Lambda$ is replaced by a multipartite (potentially entangled) state $\rho_\Lambda$.

Due to the absence of a causal link between nodes associated with different parties, we also expect any compatible distribution to respect some general linear constraints, expressing their independence in terms of the observable distribution, which are called no-signaling constraints \cite{popescu1994quantum}
 defining a set $\Nset_n$:
\begin{align}
    \nonumber
    p(a |x,y,z,\ldots) &= p(a|x,y',z',\ldots) \quad \forall a,y,y',z,z',\ldots \\
    \nonumber
    p(b |x,y,z,\ldots) &= p(b|x',y,z',\ldots) \quad \forall b,x,x',z,z',\ldots \\
    &\vdots
    \label{eq:ns_bell}
\end{align}
for each party $A,B,C,\ldots$, where $p(a|x,y,z,\ldots)$ is the marginalization of the distribution over all the other variables different from $A$.  

In the case of a single source of correlations, $\Cset(L_n), \Qset(L_n), \Nset(L_n)$ are all convex sets, and, in particular, $\Cset(L_n)$ and $\Nset(L_n)$ can be described by a finite number of constraints, making them convex polytopes. 
Moreover, it is known that $\Cset(L_n) \subset \Qset(L_n) \subset \Nset(L_n)$. 
More recently, the connection between causal modeling and Bell inequalities prompted the study of the relationships between these correlation sets in more complex causal scenarios that can also include independence between latent variables.
In this case, the geometry of the correlation sets becomes considerably more complicated, making them even non-convex in general.
In our paper, we focused on three such models, the bilocality scenario~\cite{branciard2010characterizing}, the triangle scenario ~\cite{fritz2012beyond,renou2019genuine} and the Evans Scenario~\cite{evans2016graphs,lauand2023witnessing}, which we will now describe in more detail.

% I think there's no need to define this in full generality here, is only going to be cumbersome for a causal reader and useless for someone who already knows the topic.
%---------------------------------
%If we restrict to the case of exogenous latent variables, that is, latent variables with no parents, we can give the following definition of what constitutes a quantum-compatible distribution.
%\begin{defi}[Quantum compatibility]
%
%\begin{enumerate}
%    \item Each latent variable $\Lambda \in \mL_{G}$ is associated to a quantum state $\rho_\Lambda \in L(\mH_\Lambda)$. 
%    \item Each observed variable $A \in \mO_{G}$ is represented by a POVM measurement, described by the set of operators $\left\{E_{\opa(A)}^{a}\right\}_a$, which are dependent on the outcome of its observed parents $\OPa(A)$.
%    These operators are acting non-trivially only on the space $\mH_{\LPa(A)}$ of all the latent parents of $A$.
%    Measurement operators relative to different nodes should be pairwise commuting.
%    \item The distribution is obtained by the Born rule applied to the state of all latent nodes:
%    \begin{equation}
%        p(\mO_G) = \tr \left(\prod_{a \in \mO_G} E_{\opa(A)}^{a} \; \bigotimes_{\Lambda \in \mL_{G}} \rho_\Lambda \right)
%    \end{equation}
%\end{enumerate}
%
%\end{defi}
%---------------------------------

\subsection{Bilocality scenario}
The bilocality scenario \cite{branciard2010characterizing}, represented by the DAG $B$ in Fig.~\ref{fig:bilocality_dag}, presents two independent
sources $\Lambda$ and $\Gamma$ which distribute correlations to three nodes, $A$, $B$, and $C$ that can perform measurements chosen by the settings $X$ and $Y$ for $A$ and $C$ respectively, while the central node $B$ has no external settings.
% Bilocal
As in the Bell case, we can distinguish different sets of compatible distributions associated with such a model.
The Markov condition~\eqref{eq:markov_condition} for this structure is given by
\begin{multline}
    \label{eq:bilocal}
    p(a, b, c \vert x, z) = \\ =
        \sum_{\lambda,\gamma}p(\lambda)p(\gamma)
        p(a \vert x,\lambda) p(c \vert z,\gamma)p(b\vert \lambda,\gamma) \, .
\end{multline}
As anticipated such a condition complicates considerably the characterization of the set of allowed correlation $\Cset(B)$, which is known to be non-convex, as proved by the existence of polynomial Bell inequalities~\cite{chaves2016polynomial}.

% Quantum 
Quantum distributions in this scenario $p \in \Qset(B)$ are instead defined by
 \begin{equation}
    \label{eq:quantum_bilocal}
    p(a,b,c \vert x,z) = \tr \left[ \left( A^a_x \otimes B^b \otimes C^c_z  \right) \left( \rho_\Lambda \otimes \rho_\Gamma \right) \right] \, ,
\end{equation}
for any couple of bipartite quantum states  $\rho_\Lambda, \rho_\Gamma$ and any set of POVMs with operators $A^a_x, B^b, C^c_z$.

Similarly to the multipartite Bell scenarios, here we have that some of the correlations in the quantum set $\Qset(B)$ are incompatible with a classical description~\eqref{eq:bilocal}, even when some of them are still compatible with the one in the tripartite Bell scenario $\Cset(L_3)$, showing how the assumption of independence between $\Lambda$ and $\Gamma$ increases the possibility to detect nonclassicality.

Also in this scenario, we expect correlations to respect some basic constraints given in terms of their observable distribution.
But differently from the standard multipartite Bell case, besides the linear constraints~\eqref{eq:ns_bell}, here we have some additional nonlinear ones due to the conditional independence between the two sources. More specifically, it follows that
\begin{equation}
    \label{eq:ns_bilocal}
    p(a,c \vert x,z) = \sum_b p(a,b,c \vert x,z) = p(a \vert x)p(c \vert z) \, 
\end{equation}
making this set strictly included in the set of non-signaling correlations in the tripartite Bell scenario $\Nset_3$.
We will denote this set of non-signaling bilocal correlations by $\NBset \subset \Nset_3$.

\begin{figure*}
    \centering
    \begin{subfigure}{0.3\textwidth}
    \resizebox{\textwidth}{!}{
    \begin{tikzpicture}
    	\node[var, double] (b) at (0,0) {}; %{$B_{\x\y}$};
    	\node[var, double] (a) at (4, 0) {}; %{$A_{\x\y}$};
    	\node[var, double] (c) at (-4,0) {};% {$C_{\x\y}$};
    
    	\foreach \x in {1,2} {
    	    \node[latent] (l\x) at (-2, .5+1.3*\x) {};%{$\Lambda_\x$};
    			\path[dir] (l\x) edge (c);
    			\path[dir] (l\x) edge (b);
    		
    	}
    	\foreach \y in {1,2} {
    	    \node[latent] (m\y) at (2, .5+1.3*\y) {};%{$\Mu_\y$};
    		\foreach \i in {1,2} {
    			\path[dir] (m\y) edge (b);
    			\path[dir] (m\y) edge (a);
    		}
    	}
    	\foreach \z in {1,2} {
    	    \node[latent] (n\z) at (0,-1.3*\z-.5) {};%{$\Nu_\z$};
    			\path[dir] (n\z) edge (a);
    			\path[dir] (n\z) edge (c);
    		}
    \end{tikzpicture}
    }
    \caption{2nd order quantum inflation for the Triangle scenario.}
    \end{subfigure}
    \hspace{1.5cm}
    \begin{subfigure}{0.3\textwidth}
    \resizebox{\textwidth}{!}{
    \begin{tikzpicture}
    		    \node[var, double] (b) at (0,0) {}; %{$B_{\x\y}$};
    		    \node[var, double] (a) at (4, 0) {}; %{$A_{\x\y}$};
    			\node[var, double] (c) at (-4,0) {};% {$C_{\x\y}$};
    
    	\foreach \x in {1,2,3} {
    	    \node[latent] (l\x) at (-2, .5+\x) {};%{$\Lambda_\x$};
    			\path[dir] (l\x) edge (c);
    			\path[dir] (l\x) edge (b);
    		
    	}
    	\foreach \y in {1,2,3} {
    	    \node[latent] (m\y) at (2, .5+\y) {};%{$\Mu_\y$};
    		\foreach \i in {1,2} {
    			\path[dir] (m\y) edge (b);
    			\path[dir] (m\y) edge (a);
    		}
    	}
    	\foreach \z in {1,2,3} {
    	    \node[latent] (n\z) at (0,-\z-.5) {};%{$\Nu_\z$};
    			\path[dir] (n\z) edge (a);
    			\path[dir] (n\z) edge (c);
    		}
    \end{tikzpicture}
    }
    \caption{3rd order quantum inflation for the Triangle scenario.}
    \end{subfigure}
    \\
    \begin{subfigure}{0.3\textwidth}
    \resizebox{\textwidth}{!}{
    \begin{tikzpicture}
		\node[var, double] (a) at (-3, 1) {}; %{$A_\x$};
		\node[var, double, opacity=.5] (x) at (-4, 1) {}; %{$X_\x$};
		\path[dir] (x) edge (a);

		\node[var, double] (c) at (3, 1) {};% {$C_\x$};
		\node[var, double, opacity=.5] (z) at (4, 1) {}; %{$Z_\x$};
		\path[dir] (z) edge (c);
    	
        \node[var, double] (b) at (0, 1) {}; %{$B_{\x\y}$};
    	
    	\foreach \x in {1,2} {
    	    \node[latent] (l\x) at (-1.5, 1+\x) {};%{$\Lambda_\x$};
    		\path[dir] (l\x) edge (a);
    		\path[dir] (l\x) edge (b);
    	}
    	\foreach \y in {1,2} {
    	    \node[latent] (m\y) at (1.5, 1+\y) {};%{$\Mu_\y$};
    		\path[dir] (m\y) edge (c);
    		\path[dir] (m\y) edge (b);
    	}
    \end{tikzpicture}
    }
    \caption{2nd order quantum inflation for the Bilocality scenario.}
    \end{subfigure}
    \hspace{1.5cm}
    \begin{subfigure}{0.3\textwidth}
    \resizebox{\textwidth}{!}{
    \begin{tikzpicture}
		\node[var, double] (a) at (-3, 1) {}; %{$A_\x$};
		\node[var, double, opacity=.5] (x) at (-4, 1) {}; %{$X_\x$};
		\path[dir] (x) edge (a);

		\node[var, double] (c) at (3, 1) {};% {$C_\x$};
		\node[var, double, opacity=.5] (z) at (4, 1) {}; %{$Z_\x$};
		\path[dir] (z) edge (c);
    	
    	\node[var, double] (b) at (0, 1) {}; %{$B_{\x\y}$};
    	
    	\foreach \x in {1,2,3} {
    	    \node[latent] (l\x) at (-1.5, 1+\x) {};%{$\Lambda_\x$};
    		\path[dir] (l\x) edge (a);
    		\path[dir] (l\x) edge (b);
    	}
    	\foreach \y in {1,2,3} {
    	    \node[latent] (m\y) at (1.5, 1+\y) {};%{$\Mu_\y$};
    		\path[dir] (m\y) edge (c);
    		\path[dir] (m\y) edge (b);
    	}
    \end{tikzpicture}
    }
    \caption{3rd order quantum inflation for the Bilocality scenario.}
    \end{subfigure}
    \caption{The DAGs for the orders used in our analysis for quantum inflation.
    Here we, similarly to the classical case, considered the first two levels of standard quantum inflation hierarchy~\cite{wolfe2021quantum}, generated by using $n$ copies of each latent variable. As in the other figures, latent nodes are represented by triangles while observable nodes with circles. }
    \label{fig:triangle_inflation2}
\end{figure*}
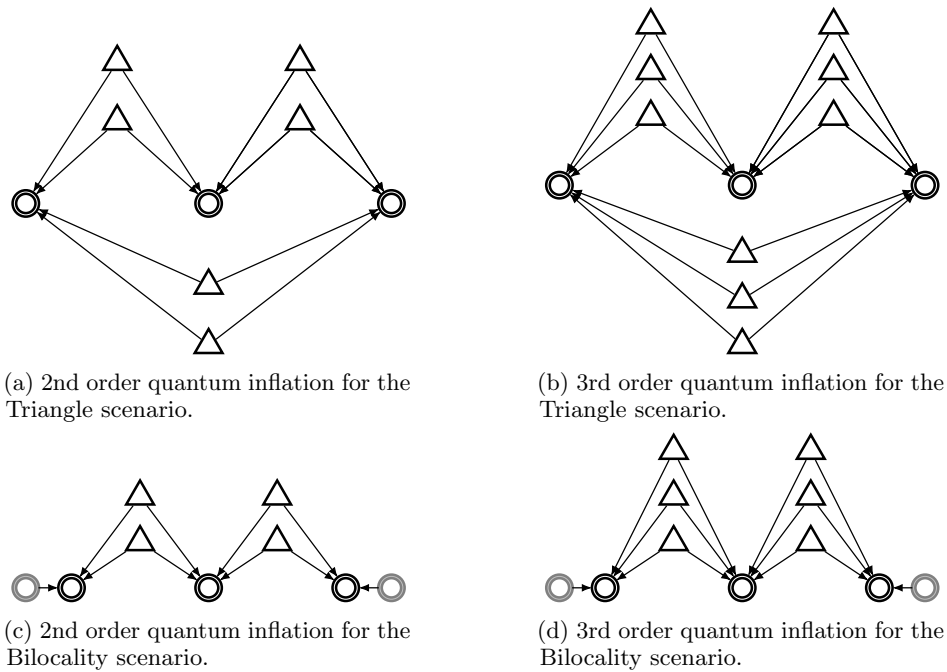

\subsection{Triangle scenario}
The triangle scenario~\cite{fritz2012beyond} is represented by the DAG in Fig.~\ref{fig:triangle_dag}, where three bipartite sources distribute systems to three separate parties forming a triangle-shaped structure. The reason why this network is of particular interest to the community is that it allows for novel non-classical phenomena~\cite{renou2019genuine,suprano2022experimental,coiteux2021no}. 
In particular, it allows for non-classicality even in the absence of measurement choices for the parties, that is, when they perform a fixed measurement~\cite{fritz2012beyond,chaves2021causal}. 

Classically, triangle correlations admit models given by
\begin{equation}
p(a,b,c)=\sum_{\lambda,\gamma,\mu}p(\lambda)p(\gamma)p(\mu)p(a| \lambda, \mu)p(b|\lambda,\gamma)p(c| \gamma,\mu),
\end{equation}
while the quantum description is given by~\footnote{Note that when evaluating $p$, one should
be attentive to which Hilbert space supports each state and
measurements.}
\begin{equation}
    p(a,b,c)=\tr \left[\left(A^a\otimes B^b \otimes C^c\right) \left( \rho_{\Lambda}\otimes\rho_{\Gamma}\otimes\rho_{M} \right)\right].
\end{equation}

Differently from the standard Bell or the bilocality cases, the triangle does not have a simple NSI description, even though a few approximations have already been proposed \cite{gisin2020constraints,henson2014theory,chaves2015information,beigi2021covariance,wolfe2019inflation}.

Furthermore, the case where all parts have binary outputs is particularly interesting. Recently,~\cite{pozaskerstjens2023postquantum} has shown that this minimal case supports post-quantum advantage, i.e. the local set and the non-signaling set do not coincide, and also that the set of triangle non-signaling correlations lies outside the quantum set. Strikingly, the conjecture that the local and quantum sets are identical~\cite{Tavakoli_2022} remains open for this minimal case. 
%Paragraph about NSI and open questions (Finner inequality vs NSI)?

\subsection{Evans scenario}

The last causal structure we will consider is the so-called Evans scenario~\cite{evans2016graphs}, which was only recently considered from a quantum perspective~\cite{lauand2023witnessing}.
The DAG $E$ for this structure is represented in Fig.~\ref{fig:evans_dag}. 
Just like in the bilocality scenario, we have three observable nodes $A, B, C$ and two latent ones $\Lambda, \Gamma$, with the difference that now there are no external inputs present, and $B$ can communicate directly its value to both $A$ and $C$.

Equation~\eqref{eq:markov_condition} in this case becomes
\begin{equation}
    \label{eq:evans}
    p(a, b, c) =
        \sum_{\lambda,\gamma}p(\lambda)p(\gamma)
        p(a \vert b, \lambda) p(c \vert b,\gamma)p(b\vert \lambda,\gamma), \, 
\end{equation}
and the quantum distribution is given by
 \begin{equation}
    \label{eq:quantum_evans}
    p(a,b,c) = \tr \left[\left( A^a_b \otimes B^b \otimes C^c_b \right) \left(\rho_\Lambda \otimes \rho_\Gamma \right)\right],
\end{equation}
for quantum states $\rho_\Lambda, \rho_\Gamma$ and POVMs $A^a_b, B^b, C^c_b$.

Despite apparent similarities with the bilocal scenario, the characterization of the classical and quantum sets $\Cset(E), \Qset(E)$ turns out to be much more complex~\cite{lauand2023witnessing}, and whether the inclusion of the former in the latter is strict or not is still an open problem. 

The set of non-signaling correlations of the Evans scenario is also poorly understood. Although a general route for deriving such theory-independent constraints for this causal structure has been proposed in \cite{lauand2023witnessing}, it relies on Fourier-Motzkin elimination which can be very costly and out of computational reach even for seemingly simple scenarios. It is known that the non-signaling set is strictly larger than the classical set of correlations $\Cset(E) \subset \Nset(E)$, at least for $|A|=3$ and $|B|=|C|=2$.

Furthermore, because there is communication between the parts, Evans's scenario allows us to go beyond passive observations of the experiment and ask what would happen if the system is intervened upon~\cite{gachechiladze2020quantifying}. We use do-conditionals $p(a|do(b))$ (and $p(c|do(b))$) to denote the probability of Alice's (Charlie's) outcome $a$ ($c$) when variable $B$ is set by force to be $b$. For classical correlations, this can be formalized with the constraint 
\begin{equation}
p(a|do(b))=\sum_{\lambda}p(\lambda)p(a|\lambda,b)
\end{equation}
and similarly for $p(c|do(b))$. Analogously, we can define what the do-conditionals would look like in terms of a quantum strategy
\begin{equation}
    p(a|do(b))= \tr \left[\left( A^a_b \otimes \mathbb{I}\right)  \rho_\Lambda \right]
\end{equation}
and the same can be done for $p(c|do(b))$.

One might be interested in exploring interventions for post-quantum theories in the Evans scenario. In order to do so, notice that the do-conditionals can be expressed in terms of a marginal probability distribution of a particular \textit{interruption} of the original graph, which consists in introducing new independent variables that each inherits one outcoming edge of the original variable, see section IX of~\cite{wolfe2021quantum} for details.
This procedure allows us to map the restrictions due to no-signaling involving do-conditionals of the Evans scenario to the no-signaling conditions on the bilocality scenario and identify the do-conditionals $p(a|do(b))$ and $p(c|do(b))$ with the marginals $p(a|x=b)$ and $p(c|z=b)$.

\subsection{Tools for detecting non-classicality}

The compatibility problem consists of answering the question: Are the statistics over the observed variables compatible with the causal structure under scrutiny? For causal structures with a single latent common cause, e.g. the Bell scenario, this amounts to a linear feasibility problem. Indeed, it is known that if we consider a decomposition of the form \ref{eq:classical_bell} we can, without loss of generality, incorporate any local randomness present in the response functions -- i.e. $p(a|x, \lambda)$, $p(b|y, \lambda)$, $p(c|z, \lambda)$ and so on -- to the source $\lambda$ and define a deterministic model. In a deterministic model, each value of $\lambda$ defines an assignment of one of the possible
outputs to each input. The model is a probabilistic mixture of these deterministic assignments of outputs to inputs, with $\lambda$ specifying which particular assignment is chosen in each run of the experiment. For each assignment, there is a corresponding \emph{local deterministic behavior} $d_{\lambda}(a,b,c..|x,y,z,..)$ and  $p(a,b,c...|x,y,z...)$ is compatible with the standard Bell scenario \emph{if, and only if} it can be expressed as a convex combination of deterministic local points. That is,
\begin{equation}\label{eq:linear_program_bell}
\begin{aligned}
    \text{\textbf{p} is local } \iff \exists &q(\lambda) \text{ s.t. } q(\lambda)\geq 0 ,\quad \sum_\lambda q(\lambda)=1\\
    &p(a,...|x,...)=\sum_{\lambda}q(\lambda)d_{\lambda}(a,...|x,...).
\end{aligned}
\end{equation}
Indeed, determining whether there exist weights $q(\lambda)$ satisfying the linear constraints in Eq. \ref{eq:linear_program_bell} is a typical instance of a \emph{linear programming problem } (LP)~\cite{BoydVandenberghe}.

For a generic causal structure, i.e. more than one source, we can use the inflation technique. Intuitively, the method works by considering the hypothetical situation where one has access to multiple copies of the sources and measurement devices that compose the network and can rearrange them in different configurations. Its core idea is to explore simple (linear) conditions of this inflated network that ultimately translate to polynomial inequalities on the observable probabilities.  It has been proven in~\cite{navascues2008convergent} the existence of a hierarchy of inflations that asymptotically converges to the classical set of correlations of any network and a test of compatibility of a given level of this hierarchy can be done via Linear Programming (LP) or Semi-Definite Programming (SDP)~\cite{BoydVandenberghe}. However, for each level $n$ of this hierarchy the memory resources required are superexponential on $n$. 
%We will represent the set associated with the $n$-th order inflation for the DAG $G$ as $\Cset_n(G)$.
Notably, this hierarchy relies on information broadcasting, a primitive that is not allowed in quantum information.

The inflation technique can constrain not only the set of classical correlations but also the set of quantum correlations a network may give rise to. \emph{Quantum inflation}~\cite{wolfe2021quantum} can be seen as a quantum analog of the classical inflation technique which avoids the latter’s reliance on information broadcasting. This is done by adapting the Navascues-Pironio-Acin (NPA) hierarchy~\cite{navascues2007bouding}, originally developed to characterize quantum correlations in Bell scenarios, in the \emph{inflated} scenario which can be tackled via noncommutative polynomial optimization (NPO) theory~\cite{navascues2008convergent}. The general goal of NPO theory is to optimize the expectation value of a polynomial over operators subject to a number of polynomial operators and statistical constraints. This optimization is achieved by means of a hierarchy of SDP tests~\footnote{These tests were implemented using~\cite{pythoninflation}.}.
%, which, as the name suggests, makes use of the aforementioned inflation technique to asses classical and quantum causal compatibility. 
The types of inflations we have used in our work are shown in Fig.~\ref{fig:triangle_inflation2}.
For each inflation level, we then study the set varying the NPA levels, and we will denote by $\Qset_{n,m}(G)$ the corresponding relaxation associated with $n$-th order inflation and level $m$ of the NPA hierarchy.
%We increase the inflation level after exhausting (i.e. pushing to the limit of the hardware, an Intel(R) Core(TM) i7-10700 CPU paired with 16Gb of RAM) the NPA levels for the current inflation.

At the core of our numerical approach is the use of quadratic programming (QP) techniques, like the \emph{branch and bound method}, that allows us to extend our optimization problems to include non-linear constraints with reasonable efficiency~\cite{gurobi}.
These techniques work by iteratively breaking the variables domain down into smaller problems that can each be approximated by a corresponding convex program. This branching subroutine enables primal and dual tasks to define upper and lower bounds that converge, up to computational precision ($10^{-9}$), to the global optimal solution.

Using this we can assess classicality for the bilocal and the Evans networks by directly imposing the independence of the sources $p(\lambda,\gamma)=p(\lambda)p(\gamma)$. 
Notice that in both scenarios we can, without loss of generality, make $\lambda$ determine the outcome $a$ for every $x$ and, similarly, $\gamma$ determine $c$ for every $z$, while $B$ has a stochastic response function $p(b|\lambda,\gamma)$. Therefore, we can take $\lambda=\{a_0,...,a_{|x|-1}\}$, $\gamma=\{c_0,...,c_{|z|-1}\}$ and $p(a,b,c|x,z)$ is bilocal \emph{if and only if}, 
\begin{equation}
    \begin{aligned}
        \exists q(\lambda,b,\gamma) \text{ s.t. }&\\
        &q\geq 0, \quad \sum_{\lambda,b\gamma} q(\lambda,b,\gamma)=1,\\
        &q(a_x=a,b,c_z=c)=p(a,b,c|x,z),\\
        &\text{ and } \quad q(\lambda,\gamma)=q(\lambda)q(\gamma).
    \end{aligned}
\end{equation}
which can be cast as a QP. For compatibility with the Evans scenario, it is sufficient to look at the same conditions but only for $x=z=b$. We also use the arguments given in~\cite{lauand2023witnessing} to extract tailored infeasibility certificates from these quadratic programs. 

\subsection{Sampling non-signaling distributions}
The starting point for the analysis of the volumes is generating the data sets to be analyzed. These points must satisfy NSI restrictions, which will be polynomial in the case of the bilocal scenario.
For this, we sample separately each coordinate of the behavior vector (or its equivalent representation by correlators) followed by a rejection step to handle the constraints~\footnote{Other ways of sampling were tested, e.g. the Monte Carlo method, which is able to include more complicated constraints to sample from. However, the rejection method was the best in terms of numerical performance; the reason for that being the cardinality of the variables and the number of constraints.}.
Specifically, we sampled points uniformly within a hypercube and then selected the subset of interest. 
We employed the results in~\cite{dyer1992volumes} to determine how many data points were necessary to sample the whole space of distributions in each situation, that is, at least $(\sim2.1)^d$ points uniformly generated in a hypercube of dimension $d$ are required to estimate a volume close to the hypercube itself.
For example, for a hypercube of dimension eight (the simplest triangle scenario in this work), one should consider data sets containing more than four hundred points to start approximating well the corresponding set of correlations. In view of such uniform distribution, the fraction of points represented by a target subset can be associated with the volume it occupies within the set containing it. More than one data set was used in what follows, therefore we are going to mention them according to their specific uses.

In all these cases, however, the idea behind them is the same: test membership of the generated points in a set of interest via LP, SDP, or QP tests. Such tests can be phrased in terms of feasibility or optimization problems. In what follows, we also investigate how these points are distributed inside the set. To this end, we consider a corresponding optimization in which the objective function to be minimized is the trace distance between the sampled distribution and some distribution inside the set (see~\cite{PhysRevA.97.022111}).

The trace distance between two probability distributions $\mathbf{p}(x)$ and $\mathbf{q}(x)$, is given by
\begin{equation}\label{eq:td}
    D(\mathbf{p},\mathbf{q}):=\frac{1}{2}\sum_{x} |p(x) -q(x)|,
\end{equation}

and the \emph{trace distance-based quantifier} $Q_{\mathcal{X}}(p)$ of the distance of a probability distribution $p(a,b,c|x,y,z)$ to a set $\mathcal{X}$ in a tripartite scenario is computed through
\begin{equation}\label{eq:quant_td}
    Q_{\mathcal{X}}(\mathbf{p})=\frac{1}{|x||y||z|} \min_{q(a,b,c|x,y,z) \in \mathcal{X}} D(p,q),
\end{equation}
and $|x|$, $|y|$, and $|z|$ stand for the cardinality of the set of measurements for Alice, Bob, and Charlie, respectively. If $\mathcal{X}$ is the local set, then we have a non-local distance, while if it is the bilocal set, we have a non-bilocal distance, and similarly to any other set. Finally, we compare the performance of the aforementioned techniques for different cases.

\section{Results}
\label{sec:results}
\subsection{Bilocality scenario}
\begin{figure}[t!]
    \centering
    \begin{tikzpicture}[line cap=round,line join=round,>=triangle 45,x=1cm,y=1cm]
\clip(-5.523418621295901,-2.6237247560632384) rectangle (2.8819419398999235,2.7521418926167223);
\fill[line width=1.2pt,dash pattern=on 2pt off 2pt,color=qqqqcc,fill=qqqqcc,fill opacity=0.45] (-4,2) -- (-4,-2) -- (0,-2) -- (0,2) -- cycle;
\fill[line width=1.5pt,color=ffbbqq,fill=ffbbqq,pattern=north east lines,pattern color=ffbbqq] (-4,-2) -- (-3.8998258956663117,-0.697431470758792) -- (-3.9204232724053525,-0.4090681964122194) -- (-4,-0.22369180576085146) -- (-4.200416331382366,-0.0037580488637113173) -- (-4,0.20885310575900728) -- (-3.9272890646516996,0.504082172351927) -- (-3.8998258956663117,0.7993112389448465) -- (-4,2) -- (-2.8012991362507953,1.8978379983603608) -- (-2.499204277411528,1.9184353750994019) -- (-2.203975210818608,2) -- (-2.00126400126302,2.212376152530682) -- (-1.8194908450231777,2) -- (-1.5173959861839106,1.904703790606708) -- (-1.2015695428519497,1.8978379983603608) -- (0,2) -- (-0.10304278343643344,0.7993112389448465) -- (-0.07557961445104554,0.5178137568446209) -- (0,0.2088531057590075) -- (0.22336113576890515,0.004732886773815095) -- (0,-0.21682601351450437) -- (-0.09617699119008646,-0.49832349561473) -- (-0.10990857568278041,-0.7866867699613025) -- (0,-2) -- (-1.2015695428519497,-1.892079321623164) -- (-1.5036644016912166,-1.9264082828548987) -- (-1.7920276760377898,-2) -- (-2.00126400126302,-2.2114013146205784) -- (-2.1971094185722615,-2) -- (-2.499204277411528,-1.905810906115858) -- (-2.8012991362507953,-1.892079321623164) -- cycle;
\fill[line width=1.5pt,color=qqqqcc,fill=qqqqcc,fill opacity=0.42] (0.6,1.1) -- (0.6,0.8) -- (0.9,0.8) -- (0.9,1.1) -- cycle;
\fill[line width=1.5pt,color=ffbbqq,fill=ffbbqq,pattern=north east lines,pattern color=ffbbqq] (0.6,0.7) -- (0.6,0.4) -- (0.9,0.4) -- (0.9,0.7) -- cycle;
\fill[line width=1.2pt,dash pattern=on 1pt off 1pt,color=qqqqcc,fill=qqqqcc,fill opacity=0.39] (0.6,1.2) -- (0.9,1.2) -- (0.9,1.5) -- (0.6,1.5) -- cycle;
\fill[line width=1.5pt,fill=none] (0.9,1.6) -- (0.9,1.9) -- (0.6,1.9) -- (0.6,1.6) -- cycle;
%\draw [line width=1pt,dash pattern=on 1pt off 1pt,color=zzccff] (-4,-2)-- (-4,2);
%\draw [line width=1pt,dash pattern=on 1pt off 1pt,color=zzccff] (-4,2)-- (0,2);
%\draw [line width=1pt,dash pattern=on 1pt off 1pt,color=zzccff] (0,2)-- (0,-2);
%\draw [line width=1pt,dash pattern=on 1pt off 1pt,color=zzccff] (0,-2)-- (-4,-2);
\draw [shift={(-15,0)},line width=1.5pt,color=qqqqcc]  plot[domain=-0.1798534997924781:0.17985349979247828,variable=\t]({1*11.180339887498947*cos(\t r)+0*11.180339887498947*sin(\t r)},{0*11.180339887498947*cos(\t r)+1*11.180339887498947*sin(\t r)});
\draw [shift={(-2,-13)},line width=1.5pt,color=qqqqcc]  plot[domain=1.3909428270024184:1.750649826587375,variable=\t]({1*11.180339887498947*cos(\t r)+0*11.180339887498947*sin(\t r)},{0*11.180339887498947*cos(\t r)+1*11.180339887498947*sin(\t r)});
\draw [shift={(11,0)},line width=1.5pt,color=qqqqcc]  plot[domain=2.961739153797315:3.3214461533822712,variable=\t]({1*11.180339887498947*cos(\t r)+0*11.180339887498947*sin(\t r)},{0*11.180339887498947*cos(\t r)+1*11.180339887498947*sin(\t r)});
\draw [shift={(-2,13)},line width=1.5pt,color=qqqqcc]  plot[domain=4.532535480592212:4.892242480177168,variable=\t]({1*11.180339887498947*cos(\t r)+0*11.180339887498947*sin(\t r)},{0*11.180339887498947*cos(\t r)+1*11.180339887498947*sin(\t r)});
\draw [shift={(-11.730485278664528,-2.083537575320275)},line width=1.5pt]  plot[domain=0.010805831824809951:0.2730339164180733,variable=\t]({1*7.730936629551389*cos(\t r)+0*7.730936629551389*sin(\t r)},{0*7.730936629551389*cos(\t r)+1*7.730936629551389*sin(\t r)});
\draw [shift={(-11.730485278664528,2.0940528898282906)},line width=1.5pt]  plot[domain=6.008980307181122:6.27101941510816,variable=\t]({1*7.731306061373008*cos(\t r)+0*7.731306061373008*sin(\t r)},{0*7.731306061373008*cos(\t r)+1*7.731306061373008*sin(\t r)});
\draw [shift={(-3.9524714521050006,-8.276632212251045)},line width=1.5pt]  plot[domain=1.2538268047676382:1.578368482991751,variable=\t]({1*6.277174623131777*cos(\t r)+0*6.277174623131777*sin(\t r)},{0*6.277174623131777*cos(\t r)+1*6.277174623131777*sin(\t r)});
\draw [shift={(-0.07720661272374864,-8.120888708067874)},line width=1.5pt]  plot[domain=1.5581833677656125:1.8898327839086744,variable=\t]({1*6.121375616444483*cos(\t r)+0*6.121375616444483*sin(\t r)},{0*6.121375616444483*cos(\t r)+1*6.121375616444483*sin(\t r)});
\draw [shift={(7.165434186087024,-2.09190052036746)},line width=1.5pt]  plot[domain=2.8462716774067847:3.12876782268071,variable=\t]({1*7.16576476695439*cos(\t r)+0*7.16576476695439*sin(\t r)},{0*7.16576476695439*cos(\t r)+1*7.16576476695439*sin(\t r)});
\draw [shift={(7.117252609683515,2.08613046205109)},line width=1.5pt]  plot[domain=3.153693707542397:3.439807352965904,variable=\t]({1*7.117773750727116*cos(\t r)+0*7.117773750727116*sin(\t r)},{0*7.117773750727116*cos(\t r)+1*7.117773750727116*sin(\t r)});
\draw [shift={(-0.04972246492671897,8.27798614416005)},line width=1.5pt]  plot[domain=4.397330793158537:4.720308944550421,variable=\t]({1*6.280745714789306*cos(\t r)+0*6.280745714789306*sin(\t r)},{0*6.280745714789306*cos(\t r)+1*6.280745714789306*sin(\t r)});
\draw [shift={(-3.9341486869069784,8.26882476156104)},line width=1.5pt]  plot[domain=4.701884796009652:5.026682872986803,variable=\t]({1*6.269170621908205*cos(\t r)+0*6.269170621908205*sin(\t r)},{0*6.269170621908205*cos(\t r)+1*6.269170621908205*sin(\t r)});
\draw [line width=1pt,dash pattern=on 3pt off 3pt,color=qqqqcc] (-4,2)-- (-4,-2);
\draw [line width=1pt,dash pattern=on 3pt off 3pt,color=qqqqcc] (-4,-2)-- (0,-2);
\draw [line width=1pt,dash pattern=on 3pt off 3pt,color=qqqqcc] (0,-2)-- (0,2);
\draw [line width=1pt,dash pattern=on 3pt off 3pt,color=qqqqcc] (0,2)-- (-4,2);
\draw [line width=1.5pt,color=ffbbqq] (-4,-2)-- (-3.8998258956663117,-0.697431470758792);
\draw [line width=1.5pt,color=ffbbqq] (-3.8998258956663117,-0.697431470758792)-- (-3.9204232724053525,-0.4090681964122194);
\draw [line width=1.5pt,color=ffbbqq] (-3.9204232724053525,-0.4090681964122194)-- (-4,-0.22369180576085146);
\draw [line width=1.5pt,color=ffbbqq] (-4,-0.22369180576085146)-- (-4.200416331382366,-0.0037580488637113173);
\draw [line width=1.5pt,color=ffbbqq] (-4.200416331382366,-0.0037580488637113173)-- (-4,0.20885310575900728);
\draw [line width=1.5pt,color=ffbbqq] (-4,0.20885310575900728)-- (-3.9272890646516996,0.504082172351927);
\draw [line width=1.5pt,color=ffbbqq] (-3.9272890646516996,0.504082172351927)-- (-3.8998258956663117,0.7993112389448465);
\draw [line width=1.5pt,color=ffbbqq] (-3.8998258956663117,0.7993112389448465)-- (-4,2);
\draw [line width=1.5pt,color=ffbbqq] (-4,2)-- (-2.8012991362507953,1.8978379983603608);
\draw [line width=1.5pt,color=ffbbqq] (-2.8012991362507953,1.8978379983603608)-- (-2.499204277411528,1.9184353750994019);
\draw [line width=1.5pt,color=ffbbqq] (-2.499204277411528,1.9184353750994019)-- (-2.203975210818608,2);
\draw [line width=1.5pt,color=ffbbqq] (-2.203975210818608,2)-- (-2.00126400126302,2.212376152530682);
\draw [line width=1.5pt,color=ffbbqq] (-2.00126400126302,2.212376152530682)-- (-1.8194908450231777,2);
\draw [line width=1.5pt,color=ffbbqq] (-1.8194908450231777,2)-- (-1.5173959861839106,1.904703790606708);
\draw [line width=1.5pt,color=ffbbqq] (-1.5173959861839106,1.904703790606708)-- (-1.2015695428519497,1.8978379983603608);
\draw [line width=1.5pt,color=ffbbqq] (-1.2015695428519497,1.8978379983603608)-- (0,2);
\draw [line width=1.5pt,color=ffbbqq] (0,2)-- (-0.10304278343643344,0.7993112389448465);
\draw [line width=1.5pt,color=ffbbqq] (-0.10304278343643344,0.7993112389448465)-- (-0.07557961445104554,0.5178137568446209);
\draw [line width=1.5pt,color=ffbbqq] (-0.07557961445104554,0.5178137568446209)-- (0,0.2088531057590075);
\draw [line width=1.5pt,color=ffbbqq] (0,0.2088531057590075)-- (0.22336113576890515,0.004732886773815095);
\draw [line width=1.5pt,color=ffbbqq] (0.22336113576890515,0.004732886773815095)-- (0,-0.21682601351450437);
\draw [line width=1.5pt,color=ffbbqq] (0,-0.21682601351450437)-- (-0.09617699119008646,-0.49832349561473);
\draw [line width=1.5pt,color=ffbbqq] (-0.09617699119008646,-0.49832349561473)-- (-0.10990857568278041,-0.7866867699613025);
\draw [line width=1.5pt,color=ffbbqq] (-0.10990857568278041,-0.7866867699613025)-- (0,-2);
\draw [line width=1.5pt,color=ffbbqq] (0,-2)-- (-1.2015695428519497,-1.892079321623164);
\draw [line width=1.5pt,color=ffbbqq] (-1.2015695428519497,-1.892079321623164)-- (-1.5036644016912166,-1.9264082828548987);
\draw [line width=1.5pt,color=ffbbqq] (-1.5036644016912166,-1.9264082828548987)-- (-1.7920276760377898,-2);
\draw [line width=1.5pt,color=ffbbqq] (-1.7920276760377898,-2)-- (-2.00126400126302,-2.2114013146205784);
\draw [line width=1.5pt,color=ffbbqq] (-2.00126400126302,-2.2114013146205784)-- (-2.1971094185722615,-2);
\draw [line width=1.5pt,color=ffbbqq] (-2.1971094185722615,-2)-- (-2.499204277411528,-1.905810906115858);
\draw [line width=1.5pt,color=ffbbqq] (-2.499204277411528,-1.905810906115858)-- (-2.8012991362507953,-1.892079321623164);
\draw [line width=1.5pt,color=ffbbqq] (-2.8012991362507953,-1.892079321623164)-- (-4,-2);
\draw (1.018209165288482,2.0541702559096553) node[anchor=north west] {$\NBset$};
\draw [line width=1.5pt,color=qqqqcc] (0.6,1.1)-- (0.6,0.8);
\draw [line width=1.5pt,color=qqqqcc] (0.6,0.8)-- (0.9,0.8);
\draw [line width=1.5pt,color=qqqqcc] (0.9,0.8)-- (0.9,1.1);
\draw [line width=1.5pt,color=qqqqcc] (0.9,1.1)-- (0.6,1.1);
\draw [line width=1.5pt,color=ffbbqq] (0.6,0.7)-- (0.6,0.4);
\draw [line width=1.5pt,color=ffbbqq] (0.6,0.4)-- (0.9,0.4);
\draw [line width=1.5pt,color=ffbbqq] (0.9,0.4)-- (0.9,0.7);
\draw [line width=1.5pt,color=ffbbqq] (0.9,0.7)-- (0.6,0.7);
\draw (1.0256343954662168,1.2373949363588326) node[anchor=north west] {$ \Cset(B)$};
\draw [line width=1.2pt,dash pattern=on 2pt off 2pt,color=qqqqcc] (0.6,1.2)-- (0.9,1.2);
\draw [line width=1.2pt,dash pattern=on 2pt off 2pt,color=qqqqcc] (0.9,1.2)-- (0.9,1.5);
\draw [line width=1.2pt,dash pattern=on 2pt off 2pt,color=qqqqcc] (0.9,1.5)-- (0.6,1.5);
\draw [line width=1.2pt,dash pattern=on 2pt off 2pt,color=qqqqcc] (0.6,1.5)-- (0.6,1.2);
\draw [line width=1.5pt] (0.9,1.6)-- (0.9,1.9);
\draw [line width=1.5pt] (0.9,1.9)-- (0.6,1.9);
\draw [line width=1.5pt] (0.6,1.9)-- (0.6,1.6);
\draw [line width=1.5pt] (0.6,1.6)-- (0.9,1.6);
\draw (1.0256343954662168,1.645782596134244) node[anchor=north west] {$\Cset(L_3)$};
\draw (1.010783935110747,0.8512829671166255) node[anchor=north west] {$\Qset(B)$};
\begin{scriptsize}
\draw [fill=rvwvcq] (-4,-2) node {};
\draw [fill=rvwvcq] (0,-2) node {};
\draw [fill=rvwvcq] (-4,2) node {};
\draw [fill=rvwvcq] (0,2) node {};
\draw [fill=rvwvcq] (-15,0) circle (2.5pt);
\draw[color=rvwvcq] (-5.48257985531836,2.867232960371611) node {$F$};
\draw [fill=rvwvcq] (-2,-13) circle (2.5pt);
\draw[color=rvwvcq] (-5.48257985531836,2.867232960371611) node {$G$};
\draw [fill=rvwvcq] (11,0) circle (2.5pt);
\draw[color=rvwvcq] (-5.48257985531836,2.867232960371611) node {$H$};
\draw [fill=rvwvcq] (-2,13) circle (2.5pt);
\draw[color=rvwvcq] (-5.48257985531836,2.867232960371611) node {$I$};
\draw [fill=rvwvcq] (-1.995948412923027,2.3063890970385583) node {};
\draw [fill=rvwvcq] (0.30988417210204267,-0.006326570329864336) node {};
\draw [fill=rvwvcq] (-1.995948412923027,-2.312159155354929) node {};
\draw [fill=rvwvcq] (-4.2880148332613786,0.0005565120134940663) node {};
\draw [fill=rvwvcq] (-11.730485278664528,-2.083537575320275) circle (2.5pt);
\draw[color=rvwvcq] (-5.48257985531836,2.867232960371611) node {$M$};
\draw [fill=rvwvcq] (-11.730485278664528,2.0940528898282906) circle (2.5pt);
\draw[color=rvwvcq] (-5.48257985531836,2.867232960371611) node {$N$};
\draw [fill=rvwvcq] (-3.9524714521050006,-8.276632212251045) circle (2.5pt);
\draw[color=rvwvcq] (-5.48257985531836,2.867232960371611) node {$O$};
\draw [fill=rvwvcq] (-0.07720661272374864,-8.120888708067874) circle (2.5pt);
\draw[color=rvwvcq] (-5.48257985531836,2.867232960371611) node {$P$};
\draw [fill=rvwvcq] (7.165434186087024,-2.09190052036746) circle (2.5pt);
\draw[color=rvwvcq] (-5.48257985531836,2.867232960371611) node {$Q$};
\draw [fill=rvwvcq] (7.117252609683515,2.08613046205109) circle (2.5pt);
\draw[color=rvwvcq] (-5.48257985531836,2.867232960371611) node {$R$};
\draw [fill=rvwvcq] (-0.04972246492671897,8.27798614416005) circle (2.5pt);
\draw[color=rvwvcq] (-5.48257985531836,2.867232960371611) node {$S$};
\draw [fill=rvwvcq] (-3.9341486869069784,8.26882476156104) circle (2.5pt);
\draw[color=rvwvcq] (-5.48257985531836,2.867232960371611) node {$T$};
\draw [fill=ffcctt] (-2,0) node {};
\draw [fill=ffcctt] (-3.8998258956663117,-0.697431470758792) node {};
\draw [fill=ffcctt] (-3.9204232724053525,-0.4090681964122194) node {};
\draw [fill=ffcctt] (-4,-0.22369180576085146) node {};
\draw [fill=ffcctt] (-4.200416331382366,-0.0037580488637113173) node {};
\draw [fill=ffcctt] (-4,0.20885310575900728) node {};
\draw [fill=ffcctt] (-3.9272890646516996,0.504082172351927) node {};
\draw [fill=ffcctt] (-3.8998258956663117,0.7993112389448465) node {};
\draw [fill=ffcctt] (-2.8012991362507953,1.8978379983603608) node {};
\draw [fill=ffcctt] (-2.499204277411528,1.9184353750994019) node {};
\draw [fill=ffcctt] (-2.203975210818608,2) node {};
\draw [fill=ffcctt] (-2.00126400126302,2.212376152530682) node {};
\draw [fill=ffcctt] (-1.8194908450231777,2) node {};
\draw [fill=ffcctt] (-1.5173959861839106,1.904703790606708) node {};
\draw [fill=ffcctt] (-1.2015695428519497,1.8978379983603608) node {};
\draw [fill=ffcctt] (-0.10304278343643344,0.7993112389448465) node {};
\draw [fill=ffcctt] (-0.07557961445104554,0.5178137568446209) node {};
\draw [fill=ffcctt] (0,0.2088531057590075) node {};
\draw [fill=ffcctt] (0.22336113576890515,0.004732886773815095) node {};
\draw [fill=ffcctt] (0,-0.21682601351450437) node {};
\draw [fill=ffcctt] (-0.09617699119008646,-0.49832349561473) node {};
\draw [fill=ffcctt] (-0.10990857568278041,-0.7866867699613025) node {};
\draw [fill=ffcctt] (-1.2015695428519497,-1.892079321623164) node {};
\draw [fill=ffcctt] (-1.5036644016912166,-1.9264082828548987) node {};
\draw [fill=ffcctt] (-1.7920276760377898,-2) node {};
\draw [fill=ffcctt] (-2.00126400126302,-2.2114013146205784) node {};
\draw [fill=ffcctt] (-2.1971094185722615,-2) node {};
\draw [fill=ffcctt] (-2.499204277411528,-1.905810906115858) node {};
\draw [fill=ffcctt] (-2.8012991362507953,-1.892079321623164) node {};
\draw [fill=qqqqcc] (0.6,1.1) node {};
\draw [fill=qqqqcc] (0.6,0.8) node {};
\draw [fill=qqqqcc] (0.9,0.8) node {};
\draw [fill=qqqqcc] (0.9,1.1) node {};
\draw [fill=ffcctt] (0.6,0.7) node {};
\draw [fill=ffcctt] (0.6,0.4) node {};
\draw [fill=ffcctt] (0.9,0.4) node {};
\draw [fill=ffcctt] (0.9,0.7) node {};
\draw [fill=rvwvcq] (0.6,1.2) node {};
\draw [fill=rvwvcq] (0.9,1.2) node {};
\draw [fill=rvwvcq] (0.9,1.5) node {};
\draw [fill=rvwvcq] (0.6,1.5) node {};
\draw [fill=rvwvcq] (0.9,1.6) node {};
\draw [fill=rvwvcq] (0.9,1.9) node {};
\draw [fill=rvwvcq] (0.6,1.9) node {};
\draw [fill=rvwvcq] (0.6,1.6) node {};
\end{scriptsize}
\end{tikzpicture}
    \caption{Illustration of the different sets of correlations in the bilocal scenario; proportions are not supposed to be faithful. Solid black lines delimit the part of the no-signalling set compatible with the bilocal conditions, while the blue region with a dashed border determines the local set and solid blue lines delimit the bilocal subset within it. The hashed yellow region encompasses quantum behaviors.}
    \label{fig:sets_bilocality}
\end{figure}
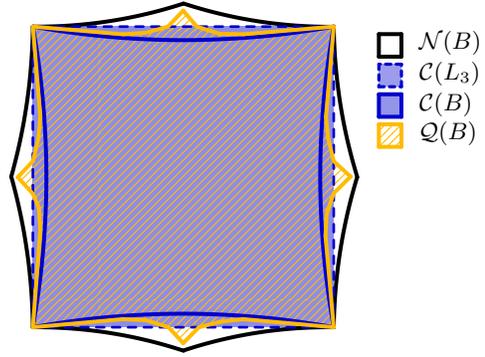

In the case of the bilocality scenario, we estimated the relative volume among the different sets of correlation $\mathcal{C}(B)$, $\mathcal{Q}(B)$ and $\mathcal{C}(L_3)$ with respect to randomly sampled points inside $\NBset$ for the case where all variables are binary. 
Fig.~\ref{fig:sets_bilocality} schematically illustrates these sets.

For doing so, we parametrize the conditional probability distribution $p(a, b, c|x, z)$ in terms of the single-, two- and three-party correlators $\langle A_x \rangle$, $\langle B \rangle$, $\langle C_z \rangle$, $\langle A_x B\rangle$, $\langle BC_z \rangle$,  $\langle A_xC_z \rangle$ =  $\langle A_x \rangle$$\langle C_z \rangle$ and  $\langle A_xBC_z \rangle$ for all $x,z\in \{0,1\}$, where $\langle A_xBC_z \rangle=\sum_{a,b,c}(-1)^{a+b+c}p(a,b,c|x,z)$, and similiarly for the other correlators. This yields a total of 13 parameters in the interval $[-1, 1]$. 
We generated a set of uniformly distributed points in the 13-dimensional hypercube and, naturally, considered only the ones inside the region of non-negative probability distributions, i.e. we excluded the points that do not respect 
\begin{eqnarray}
    & & p(a,b,c|x,z) = \\ \nonumber
    & &\dfrac{1}{8} \Big( 1+(-1)^a\langle A_x \rangle+(-1)^b\langle B \rangle+(-1)^c\langle C_z \rangle+\\ \nonumber
    & & +(-1)^{a+b}\langle A_x B\rangle + (-1)^{b+c}\langle BC_z \rangle +\\ \nonumber
    & & +(-1)^{a+c}\langle A_x \rangle\langle C_z \rangle +(-1)^{a+b+c}\langle A_xBC_z \rangle  \Big) \geq 0,
\end{eqnarray}
which gives us a set of uniformly distributed points in $\NBset$. 
We remark that this can be done without loss of generality for all non-signaling behaviors, i.e. one can always choose some minimal representation or (for the case of binary outcomes) correlators, i.e. expectation values, representation to isolate the free parameters and eliminate all equality constraints and, if necessary, use inequality constraints to filter valid probability distributions from the data set that are uniformly distributed in the region under scrutiny. 

First, we analyze which points in our data set violate the standard tripartite Bell locality by computing their trace distance to the local set of distributions $\Cset(L_3)$. 
Then, for the remaining points, which have an explicit local decomposition, we analyze different levels of the classical inflation hierarchy $\Cset_n(B)$ to compute their distances from the bilocal set $\Cset(B)$, obtained by solving directly the QP problem, and, from them, estimate the relative volumes of these sets. The results for the volumes are presented in Table~\ref{tab:bilocal_volumes}. 
Interestingly, while the set of non-local distributions $\comp \Cset(L_3) = \NBset \setminus \Cset(L_3)$ amounts to approximately only $2.39 \%$, the non-bilocal set $\comp \Cset(B) = \NBset \setminus \Cset(B)$ occupies $32.6 \%$ of the correlation space, a clear signature of the advantage of considering the independence of the sources when testing the non-classicality of the data.

%\begin{figure}[t!]
%    \centering
%    \includegraphics[width=0.45\textwidth]{set_bilocality_new.png}
%    \caption{Illustration of the different sets of correlations in the bilocal scenario; proportions are not supposed to be faithful. Solid black lines delimit the part of the no-signalling set compatible with the bilocal conditions, while the blue region with a dashed border determines the local set and solid blue lines delimit the bilocal subset within it. The hashed yellow region encompasses quantum behaviors.}
%    \label{fig:sets_bilocality}
%\end{figure}

%figure above using tikz latex code

Regarding the use of inflation, the best witness of non-classicality is represented by $\Cset_3$, i.e. $3$rd order inflation, showing that $13.3\%$ of its volume is non-bilocal, still a significantly smaller value than that obtained by QP. 
Besides being less accurate in detecting non-classicality, the inflation technique is computationally more demanding (around fifty times more as compared with the QP approach), which is the reason why we analyzed two independently generated samples of different sizes: $N=10^4$ for the inflation and $N=10^6$ in the QP case.

\begin{table}[ht]
    \centering
    \begin{tabular}{lr}
         %\hline
         Set & Volume\\
         \hline
         $\comp \Cset(L_3)$ (Non-local) & 0.0239\\
         $\comp \Cset(B)$ (Non-bilocal) & 0.326\\
         $\comp \Cset_1(B)$ (Inflation 1st order) & 0.0233\\
         %$\comp \Cset_2(B)$ (Inflation 2nd order) & 0.187\textcolor{red}{$\rightarrow0.0999$}\\
         %$\comp \Cset_3(B)$ (Inflation 3rd order) & 0.013\textcolor{red}{$\rightarrow0.133$}
         $\comp \Cset_2(B)$ (Inflation 2nd order) & 0.0999\\
         $\comp \Cset_3(B)$ (Inflation 3rd order) & 0.133
    \end{tabular}
    \caption{Volumes of different correlation sets in the bilocality scenario. In the first two rows we have the results provided by QP on a data set with $10^6$ samples, while in the last three, we have the volume associated with the set of non-bilocal behaviors $\comp \Cset_n(B)$ computed for different orders of inflation $n$, on a data set with only $10^4$ instances. Order $n = 3$ was the highest configuration analyzed. Which is still considerably far from the value obtained by the QP approach.}
    \label{tab:bilocal_volumes}
\end{table}
%Additionally, we estimate the volumes of different levels of the quantum inflation hierarchy \textcolor{red}{(and scalar extension ?)}\cite{quantum_inflation} which, although are outer approximations of the quantum set of correlations, have been proven to be asymptotically convergent for the bilocal scenario \cite{convergent_DGross}. \textcolor{red}{[need to put references]} We give the explicit estimations in Table {\color{red} (Table with quantum volumes here)}

Additionally, considering $\Qset_{2,3}$, i.e. $2$nd order of inflation together with the $3$rd level of the NPA hierarchy, we estimated the ratio of the quantum volume with respect to the other sets. 
As can be seen in Table~\ref{tab:qvolumes_bilocal}, while the quantum volume (with the NPA level considered) within the non-local correlations is $43 \%$, it increases to about $86 \%$ when considering the set of non-bilocal correlations. 
 
\begin{table}[ht]
    \begin{tabular}{lr}
        %\hline
        Set & Volume of $\Qset_{2,3}(B)$\\
        \hline
        %$\NBset$ (Non-signaling) & 0.77\textcolor{red}{$\rightarrow0.952$}\\
        $\NBset$ (Non-signaling) & 0.952\\
        $\comp \Cset(L_3)$ (Non-local) & 0.43\\
        %$\Cset(L_3)$ Local & 0.76\textcolor{red}{$\rightarrow0.942$}\\
        $\Cset(L_3)$ Local & 0.942\\
        %$\comp \Cset(B)$ (Non-bilocal) & 0.092\textcolor{red}{$\rightarrow0.856$}\\
        $\comp \Cset(B)$ (Non-bilocal) & 0.856\\
        $\Cset(B)$ (Bilocal) & 1.0\\
    \end{tabular}
    \caption{Volumes of the behaviors that are compatible with a quantum description, evaluated considering the relaxation $\Qset_{2,3}(B)$, i.e. the $2$nd order of quantum inflation and NPA level $3$. The results were obtained with a data set containing $10^4$ behaviors.}
    \label{tab:qvolumes_bilocal}
\end{table}

\begin{figure*}[t!]
    \centering
    \begin{subfigure}[b]{0.45\textwidth}
    \includegraphics[width=1\textwidth]{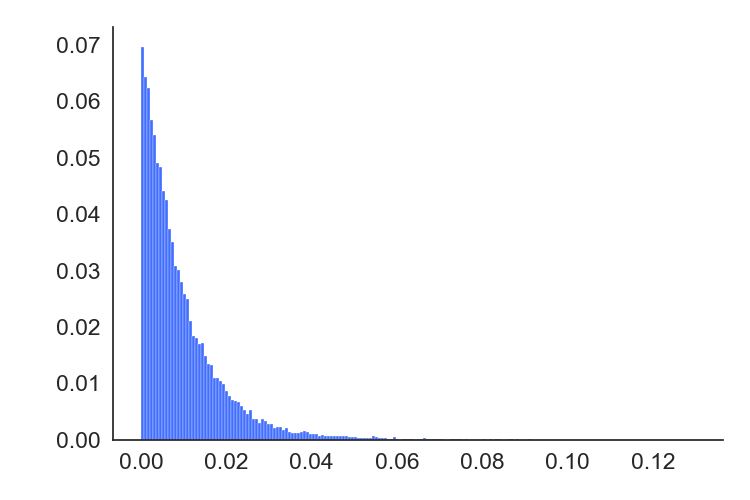}
    \caption{Non-local distances}
    \label{subfig:nl_dist}
    \end{subfigure}
    \hspace{1cm}
    \begin{subfigure}[b]{0.45\textwidth}
    \centering
    \includegraphics[width=1\textwidth]{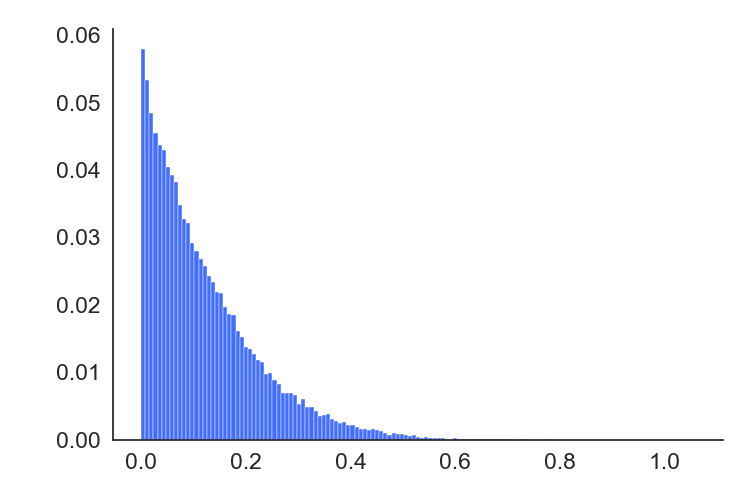}
    \caption{Non-bilocal set.}
    \label{subfig:nbl_dist_all}
    \end{subfigure}
    
    \begin{subfigure}[b]{0.45\textwidth}
    \centering
    \includegraphics[width=1\textwidth]{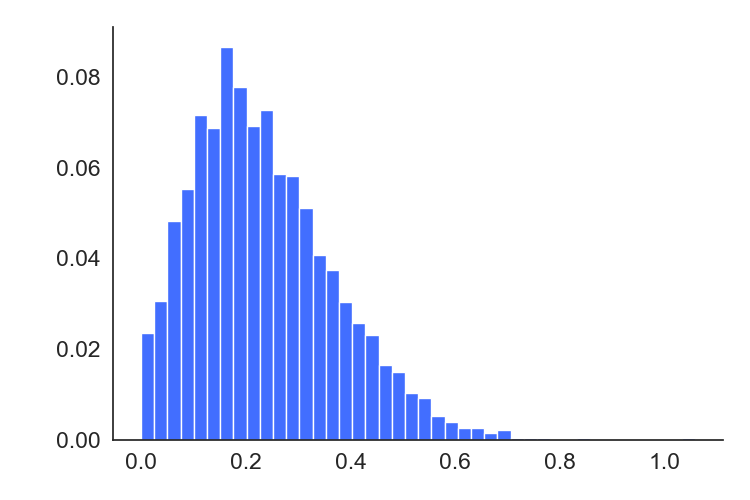}
    \caption{Non-bilocal within non-local set.}
    \label{subfig:nbl_dist_nl}
    \end{subfigure}
    \hspace{1cm}
    \begin{subfigure}[b]{0.45\textwidth}
    \centering
    \includegraphics[width=1\textwidth]{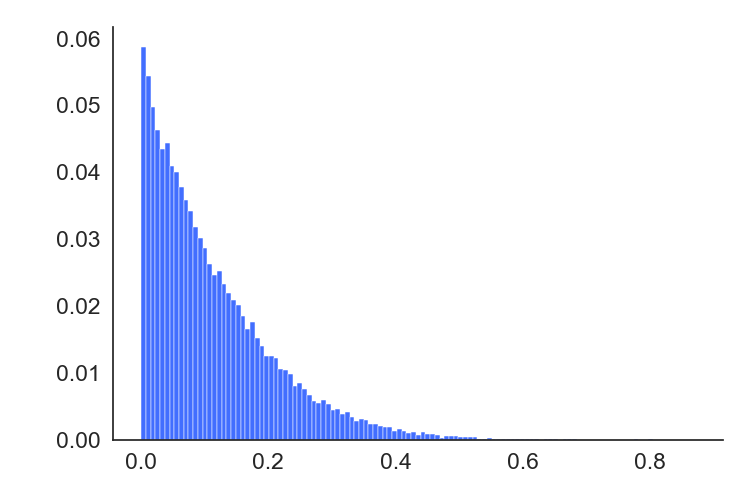}
    \caption{Non-bilocal within local set.}
    \label{subfig:nbl_dist_l_nbl}
    \end{subfigure}
    \caption{Distributions of distances in the bilocal scenario. In~\textbf{(\subref{subfig:nl_dist})} we have non-local distances for around $2 \times 10^4$ behaviors with a gap precision between primal and dual LP tasks equal to 0\%.
    In~\textbf{(\subref{subfig:nbl_dist_all})} one can see how non-bilocal distances are distributed considering points within the whole non-bilocal set, while in (c) and (d) we split such set between non-local and local, respectively, to offer a different perspective and highlight the concentration observed in (c); 
    results for around $4 \times 10^4$ behaviors using precision of 10\% for the gap between primal and dual solutions in this case.}
    \label{fig:nl_nbl_dist}
\end{figure*}

To gather more information about the structure of different correlation sets, beyond their relative volumes, we can also estimate how these points are distributed relative to their trace distances inside $\NBset$. From Fig.~\ref{fig:nl_nbl_dist} we can see how the points in $\NBset$ have their distances from the local $\Cset(L_3)$ and bilocal $\Cset(B)$ set distributed according to a Poissonian-like distribution. An interesting observation here is the presence of a concentration of the distances for behaviors that are non-classical, i.e. the distribution is peaked at a small distance from the local and bilocal sets, which can be seen as an instance of the concentration phenomena reported in~\cite{duarte2018concentration}.

\subsection{Triangle scenario}
Differently from the bilocality and Evans scenarios, for which there is a natural set of NSI correlations to sample from, in the triangle scenario there is no simple way to enforce the no-signaling and independence conditions~\cite{gisin2020constraints}. For this reason, we analyze different volumes of sets of correlations relative to points sampled inside the 8-dimensional simplex, that is, sampling over all well-defined probability distributions $p(a,b,c)$ with $a,b,c=0,1$ being dichotomic variables. Also, given that the triangle scenario imposes non-quadratic constraints there is no direct manner to use quadratic optimizers, for this reason we rely on the inflation technique~\cite{wolfe2019inflation, wolfe2021quantum} as well as the covariance approach~\cite{kela2019semidefinite,aaberg2020semidefinite} and the Finner~\cite{renou2019limits} and Shannon-type \cite{fritz2012entropic,chaves2015information} inequalities.

First, we compare the set of classical and quantum correlations using the inflation technique on our data points. Table~\ref{subtab:tri_cq_inflations} shows the results for different orders of classical inflations $\Cset_m(T)$ with $m=1,2,3$, and for different quantum inflations $\Qset_{m,n}(T)$ with $m=1,2$ and up to level $n=5$ of the NPA hierarchy.
Notice how we are unable to detect non-quantum behaviors without considering one extra copy of each source at least. Moreover, we have found points that are incompatible with a classical description but are quantum-compatible with $\Qset_{2,4}$, i.e. up to the $2nd$ order inflation and NPA level $4$. These points offer good candidates to resolve a still open question: whether quantum non-classical correlations can emerge in the triangle scenario with all variables dichotomic~\cite{pozas2023post}.

%\textcolor{red}{[Maybe it is a good idea to show a figure with the inflations we are considering in each case and explain in a bit more detail what we mean with each level of the NPA hierarchy. This could be done in an Appendix if needed.]}

Then, we consider the covariance decomposition test --- which can only test the network topology and as such are valid for all GPTs~\cite{beigi2021covariance} --- and the volume delimited by the Finner and the Shannon-type entropic inequalities found in \cite{fritz2012entropic,chaves2015information}; the first of them are proven to follow for all quantum distributions and conjectured to be satisfied for all non-signaling distributions~\cite{renou2019limits} while the entropic ones, as the covariance test, tell us about the topology only. 
Table~\ref{subtab:tri_other_methods} summarizes these results. Among them, we can see that the covariance test managed to exclude the largest number of behaviors.
%It should be pointed out, however, that only a small number of entropic inequalities for this scenario have been used for this comparison, because determining all of them is computationally too demanding, as mentioned in the reference.} Já adicionei a info que estamos falando de desigualdades Shannon type

\begin{table}[ht]
    \centering
    \begin{subtable}{0.5\textwidth}
        \begin{tabular}{lr}
        Method & Incompatible volume\\
        \hline
        Covariance & 0.0369\\
        Finner & 0.0191\\
        Entropic & 0.00095\\
    \end{tabular}
    \caption{Comparison of other methods to detect incompatibility with the triangle network.}
    \label{subtab:tri_other_methods}
    \end{subtable}
    %\begin{subtable}{0.5\textwidth}
    %    \begin{tabular}{|c|c|c|}
    %    \hline
    %     Inflation level & NPA level & Incompatible\\
    %     \hline
    %     1 & 3 & 0.0 \\
    %     \hline
    %     2 & 2 & 0.08856\\
    %     \hline
    %     2 & 3 & 0.1103\\
    %     \hline
    %     3 & 2 & 0.1016\\
    %     \hline
    %    \end{tabular}
    %\caption{Volumes of behaviours incompatible with a classical description.}
    %\label{subtab:tri_classical_inflations}
    %\end{subtable}
    %\begin{subtable}{0.5\textwidth}
    %    \begin{tabular}{|c|c|c|}
    %         \hline
    %         Inflation level & NPA level & Incompatible \\
    %         \hline
    %         1 & 5 & 0.0 \\
    %         \hline
    %         2 & 3 & 0.1098\\
    %         \hline
    %    \end{tabular}    %\begin{subtable}{0.5\textwidth}
    %    \begin{tabular}{|c|c|c|}
    %    \hline
    %     Inflation level & NPA level & Incompatible\\
    %     \hline
    %     1 & 3 & 0.0 \\
    %     \hline
    %     2 & 2 & 0.08856\\
    %     \hline
    %     2 & 3 & 0.1103\\
    %     \hline
    %     3 & 2 & 0.1016\\
    %     \hline
    %    \end{tabular}
    %\caption{Volumes of behaviours incompatible with a classical description.}
    %\label{subtab:tri_classical_inflations}
    %\end{subtable}
    \begin{subtable}{0.5\textwidth}
        \begin{tabular}{lr}
             \hline
             Set & Volume \\
             \hline
             $\comp \Cset_1(T)$ & 0.0 \\
             %$\comp \Cset_2(T)$ & 0.1103\textcolor{red}{$\rightarrow0.09592$} \\
             $\comp \Cset_2(T)$ & 0.09592 \\
             %$\comp \Cset_3(T)$ & 0.1016\textcolor{red}{(old)} \\
             $\comp \Cset_3(T)$ & 0.113* \\
             $\comp \Qset_{1,5}(T)$ & 0.0 \\
             $\comp \Qset_{2,3}(T)$ & 0.1098\\
        \end{tabular}
    \caption{Volumes of behaviors incompatible with a classical or quantum description. *Here we use about 10\% of the data set due to the computational cost.}
    \label{subtab:tri_cq_inflations}
    \end{subtable}
    \caption{\textbf{(\subref{subtab:tri_other_methods})} Volume of incompatible distributions calculated with other known methods, specifically the Finner and the entropic inequalities~\cite{fritz2012entropic} and the covariance method~\cite{beigi2021covariance}. \textbf{(\subref{subtab:tri_cq_inflations})} Volumes of points incompatible with classical $\Cset_n(T)$ and quantum inflations $\Qset_{m,n}(T)$ for the triangle scenario using different configurations of $n,m$, on a data set of around $10^5$ instances.}
    \label{tab:triangle_volumes}
\end{table}

\subsection{Evans scenario}
Now we move our attention to the Evans scenario. We start by pointing out that the relationship between the Evans and the bilocality scenario, which was first noticed for classical and quantum correlations~\cite{lauand2023witnessing}, indeed holds to all GPTs. In fact, if one has access to GPT states and measurements one can uniquely define a mapping between Evans and the bilocal scenario by making an identification of the measurement effects of the variables $A$ and $C$ in each scenario, analogously to what was argued in~\cite{lauand2023witnessing}. 
Therefore, the probability distribution $p(a,b,c)$ in the Evans scenario is compatible with \emph{any} GPT \emph{if and only if} there exists a bilocal distribution $p_B(a,b,c|x,z)$ compatible with the same GPT which satisfies $p_B(a,b,c|x=z=b)=p(a,b,c)$. 
Using this mapping, we can explore the set of non-signaling correlations in the Evans scenario by virtue of the non-signaling conditions of the bilocality scenario.

We can formalize this with the following statement:
\begin{equation}
\begin{aligned}
\label{eq: NSI_Evans}
    &p(a,b,c) \text{ is NSI-compatible} \iff \\
    &\exists p_B(a,b,c|x,z) \text{non-signaling distribution,}\\
    & \text{ s.t. } p_B(a,b,c|x=z=b)=p(a,b,c) \\
    &\quad \quad \quad\text{ and } p_B(a,c|x,z)=p_B(a|x)p_B(c|z).
\end{aligned}
\end{equation}

This tells us how to explore the set of NSI-compatible correlations in the Evans scenario using a single quadratic program. Furthermore, this fact allows our test to be conclusive, i.e. if one can prove that the conditions from~\ref{eq: NSI_Evans} do not hold for some candidate distribution $p(a,b,c)$ then we can conclude the distribution is NSI-incompatible and, conversely, if one can find a solution then there exists a no-signaling distribution $p_B(a,b,c|x,z)$ that respects the NSI requirements, such that recovers $p(a,b,c)$ by setting $x=z=b$, leaving no ambiguity up to a small ($\sim 10^{-9}$) computational precision. 
We remark that this is not the case for the inflation tests, as they constitute only necessary conditions for compatibility. 

\begin{figure*}[t!]
    \centering
    \begin{subfigure}[b]{0.45\textwidth}
    \centering
    \includegraphics[width=1\textwidth]{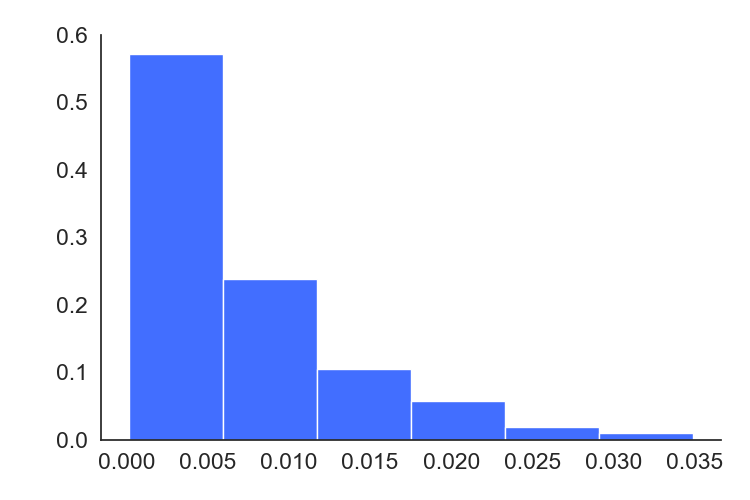}
    \caption{Non-Evans with passive observations.}
    \label{subfig:ne_wo_interv_dist}
    \end{subfigure}
    \hspace{1cm}
    \begin{subfigure}[b]{0.45\textwidth}
    \centering
    \includegraphics[width=1\textwidth]{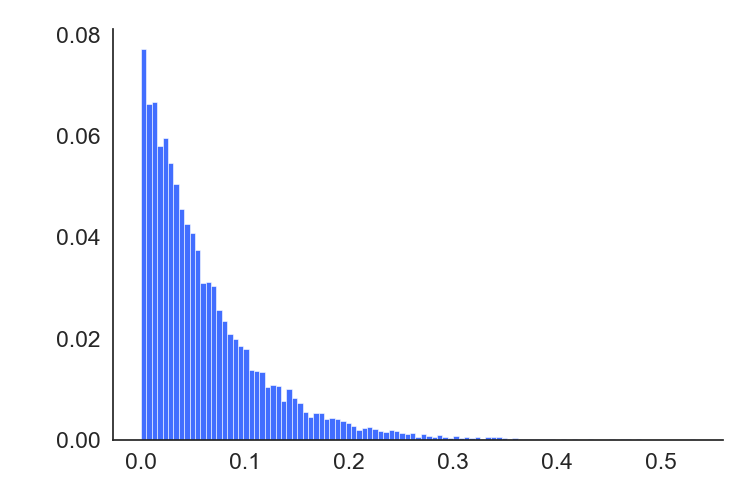}
    \caption{Non-Evans under intervention.}
    \label{subfig:ne_interv_dist}
    \end{subfigure}
    \caption{The distribution of the trace-distances of non-classical behaviors to the set of those compatible with a classical description when (\subref{subfig:ne_wo_interv_dist}) only passive 
    observations are made and (\subref{subfig:ne_interv_dist}) interventional data is also taken into account. While in (\subref{subfig:ne_wo_interv_dist}) all the non-classical instances have been used in the plot, in (\subref{subfig:ne_interv_dist}) only $10\%$ has been considered.}% The optimizations have been run to complete optimality (i.e. no gap between primal and dual tasks).}
    \label{fig:ne_dist}
\end{figure*}

We can, thus, estimate the volume of the Evans non-signaling set $\Nset(E)$ inside the simplex of probability distributions, see Table~\ref{tab:NSI_estimate}, which includes also results using different numbers of outputs for each party. 
Naively, one might believe that non-signaling is not relevant in the presence of signals between parties of the network and in the absence of inputs, but this is indeed not true as we can see that, approximately, the non-signaling set represents $84.94\%$ of the volume relative to the simplex of all probability distributions. Notice also how the volume increases as we increase the number of outputs on Bob's side, whilst a similar change on the other parties causes the opposite effect.
\begin{table}[ht]
    \centering
    \begin{tabular}{cccr}
         $|A|$ & $|B|$ & $|C|$ & Volume\\
         \hline
         2 & 2 & 2 & 0.8494 \\
         2 & 3 & 2 & 0.9823 \\
         3 & 2 & 3 & 0.7475\\
    \end{tabular}
    \caption{Volume of the non-signaling set $\Nset(E)$ within the simplex of valid probability distributions in the Evans scenario for different numbers of outputs for each party. The number of samples considered in each case was of the order of $10^5$.}
    \label{tab:NSI_estimate}
\end{table}

Naturally, we also investigated the volume of the classically compatible distribution in the Evans scenario $\Cset(E)$ and, remarkably, found a very small gap ($\approx 0.12\% $) of NSI-compatible non-classical distributions in the minimal case where all variables are bits. 
Fig.~\ref{subfig:ne_wo_interv_dist} shows the distribution of the distances to the classical set. 
All non-classical distributions that we have found cannot be ruled out by quantum inflation up to $2nd$ order inflation and NPA level $3$, therefore we cannot tell that these points are truly post-quantum with our current techniques. 
Even the classical version of the inflation technique is unable to exclude any of the non-classical points we found. 
This opens the interesting possibility (with actual candidate probability distributions) that a classical-quantum gap exists in the Evans scenario, a question that remains open~\cite{lauand2023witnessing}.

In particular, starting from these points, we can propose the candidate distribution
\begin{equation}
    \begin{aligned}
        &P_{NS}(1,0,0)=2/81, \quad P_{NS}(0,0,1)=1/55\\
        &P_{NS}(0,1,0)=1/11,\quad P_{NS}(1,0,1)=1/5\\
        &P_{NS}(1,1,0)=P_{NS}(0,1,1)=1/81,\\
        &P_{NS}(1,1,1)=1/2\sqrt{2},\\
        &P_{NS}(0,0,0)=1-\sum_{a,b,c\neq0,0,0}P_{NS}(a,b,c)\\
    \end{aligned}
\end{equation}
which satisfies the NSI test~\ref{eq: NSI_Evans} and can be certified to be non-classical with QP with a corresponding witness given by
\begin{equation}
   W:=\sum_{a,b,c}( p(a,b,c)-P_{NS}(a,b,c))^2\geq \frac{1}{36.853}.
\end{equation}

Moreover, we analyzed what happens if one considers additional interventional data $p(a,c|do(b))$. For the particular case of the Evans causal structure, it is sufficient to provide only $p(a|do(b))$ and $p(c|do(b))$, since $p(a,c|do(b))=p(a|do(b))p(c|do(b))$. To do so, we use the interruption technique to map valid non-signaling probability distributions in the bilocal scenario to valid hybrid data tables in the Evans scenario. This is done by considering only $p_{B}(a,b,c|x=z=b)$ and the marginals $p_B(a|x)$,$p_B(c|z)$ and identifying them with $p(a,b,c)$, $p(a|do(b=x))$ and $p(c|do(b=z))$ respectively. 

Similar to the observable case, we can ask what portion of these valid non-signaling hybrid data tables are classically achievable. Remarkably, we found that interventions increase the power to detect non-classicality by two orders of magnitude. Indeed, the non-classical volume, in the case where all variables are binary, increases to $14.6\%$ relative to the number of non-signaling distributions sampled, as opposed to $0.12\%$ using only passive observations. We also look at how these points are distributed relative to their trace distances inside the NSI set, see Fig.~\ref{subfig:ne_interv_dist}.

\section{Discussion}
\label{sec:discussion}

From a modern viewpoint, Bell's theorem can be seen as an instance of a causal inference problem, more precisely causal compatibility where we impose a given causal structure on a quantum experiment and ask whether a classical causal model can explain the observed correlations. This simple yet powerful realization led to a number of generalizations of Bell's theorem for causal networks of growing size and complexity, showing, for instance, the emergence of non-classical behavior even without the need for measurement choices~\cite{fritz2012beyond,renou2019genuine,polino2023experimental,chaves2021causal} or by allowing time-like rather than space-like correlation scenarios~\cite{chaves2018quantum}. 
Of particular relevance, are causal networks composed of independent sources of correlations, scenarios that unveiled new features such as the possibility of self-testing all entangled quantum states~\cite{vsupic2023quantum} and quantum theory itself~\cite{weilenmann2020self}, activation of non-classical behaviour~\cite{pozas2019bounding,poderini2020criteria}, refined notions of multipartite non-classicality~\cite{pozas2022full,suprano2022experimental} and novel tests of the role of complex numbers in quantum mechanics~\cite{renou2021quantum}.

A basic problem within this context is that of characterizing the sets of correlations allowed by each causal structure according to classical, quantum, and non-signaling theories. A problem that, differently from the standard Bell scenario, relies on polynomial causal constraints that impose a non-convex structure to the set of allowed probability distributions compatible with a given causal network. Recently a number of different tools have been proposed to approach this problem but even simple causal networks still have a very fragmented and partial characterization. Moreover, it is unclear how effective those different methods are in witnessing non-classicality. To obtain a more coherent and global picture of both the sets of correlations as well as the tools designed to address them, we analyzed the volumes of such different sets of correlations.

Considering the simplest bilocality scenario and using QP we obtained that only $2.4 \%$ of the NSI correlations are non-local while $32.6 \%$ are non-bilocal, thus showing that the ability to witness non-classicality is significantly enhanced if we take into account the independence of the sources. Furthermore, the distribution of distances of the non-classical points to the local and bilocal sets show an exponential decay, meaning that most of them are concentrated close to the classical sets. In comparison, the best results we obtained with the inflation technique -- corresponding to a $3$rd order inflation level 
% in the $3$rd level of the NPA hierarchy 
-- provide a lower bound of  $13.3\%$ for the volume of the non-bilocal set, meaning that more than half of the non-bilocal points are not detected by this method.

In the Evans scenario, we have shown that NSI correlations occupy a significant volume of the simplex set (the set of all probability distributions), in some cases surpassing $98\%$. Surprisingly, only $0.12 \%$ of those NSI correlations are actually non-classical. In comparison, the inflation technique again shows relatively poor performance, being unable to detect any of these non-classical points (up to the level of the hierarchy we could handle numerically). Remarkably, however, the volume of non-classical correlations is increased to $14.6 \%$ when we consider also the effect of interventions in the Evans scenario, a clear signature that interventional data can enhance significantly our ability to witness non-classicality.

For the triangle network, since it involves third-order polynomial constraints, we cannot directly use QP and, for this reason, we have employed and compared four specific tools: the inflation technique, the covariance test, the Finner inequality, and Shannon-type entropic inequalities. Once more, the $3$rd inflation level was the best approximation we achieved with the inflation technique, lower bounding the volume of non-classical correlations to $11.3^*\%$ of the total set of tripartite probability distributions. Interestingly, the lower bound on the volume of post-quantum correlations is $10.98 \%$, pointing out that only a small fraction of the non-classical correlations in the triangle scenario might have a quantum description. In comparison, the Finner inequality detects a volume of $1.91 \%$ of non-classical correlations while the covariance approach and the entropic inequalities lead to $3.69\%$ and $0.095\%$, respectively.

In summary, our approach unveils a few interesting features of the non-convex sets of correlations of causal networks of relevance in the literature. On the positive side, it shows that taking into account the independence of the sources as well as interventional data can greatly improve the volume of non-classical correlations, enhancing our ability to witness them. In turn, the inflation technique, the most general tool at the disposal, cannot detect a significant portion of non-classical correlations, at least the approximation level that was computationally accessible.

This shows that new tools might be needed to advance our understanding of such networks and the non-classical features they entail. One interesting possibility is to adapt the inflation method to generate quadratic constraints only, which could then be efficiently handled by a quadratic optimizer. Another relevant direction is to use these results as the starting point to solve open questions. For instance, we have detected non-classical points in the simplest Evans and triangle scenarios that nonetheless pass the test of quantum inflation. Those are good candidates for possible quantum violations of the causal constraints imposed by such networks and we hope our work motivates further research in those directions.

\section*{Acknowledgements}
The authors thank the International Institute of Physics for hosting the workshop ´´Quantum Causality Retreat'', during which this work has started to be discussed. This work was supported by the Serrapilheira Institute (Grant No. Serra-1708-15763), the Simons Foundation (Grant Number 1023171, RC), the Brazilian National Council for Scientific and Technological Development (CNPq) (INCT-IQ and Grant No 307295/2020-6), the São Paulo Research Foundation FAPESP (Grant Nos. 2018/07258-7, 2022/03792-4) and the Brazilian agencies MCTIC, CAPES and MEC.

\section{Code availability}
We provide the codes to reproduce our results at the link \url{https://github.com/Giuhcs/relative_volumes_tripartite}.

%%%% remove this command at the end %%%%
\clearpage

\bibliography{main}
\end{document}